\documentclass[%
 reprint,
amsmath,amssymb,
aps,prx, twocolumn, superscriptaddress, showpacs
]{revtex4-2}

\usepackage{amsthm}
\renewcommand\qedsymbol{$\blacksquare$}

\usepackage{graphicx}
\usepackage{dcolumn}
\usepackage{bm}
\usepackage{enumitem}

\usepackage{booktabs}
\usepackage{setspace} 
\usepackage{algorithm}
\usepackage{algpseudocode}
\usepackage{xcolor}
\usepackage{tikz}
\usepackage{circuitikz}
\usetikzlibrary{decorations.pathreplacing}  
\usepackage{multirow,array}

\usepackage{adjustbox,tabularx}
\usepackage{subcaption}   

\usepackage{hyperref}
\hypersetup{
    colorlinks=true,
    linkcolor=blue,
    citecolor=blue,
    filecolor=blue,
    urlcolor=blue
}

\def\>{\ensuremath{\rangle}}
\def\<{\ensuremath{\langle}}

\newcommand {\Tr} {{\mathrm{Tr}}}

\newcommand{\bra}[1]{\langle #1 \vert}

\newcommand{\ket}[1]{|#1\rangle}

\newtheorem{res}{Result}[section]

\newtheorem{lem}{Lemma}[section]

\begin{document}

\preprint{APS/123-QED}

\title{Realization of Thread Level Parallelism on Quantum Devices}

\author{Keren Li}
\email{likr@szu.edu.cn}
\affiliation{College of Physics and Optoelectronic Engineering, Shenzhen University, Shenzhen 518060, China}
\affiliation{Quantum Science Center of Guangdong-Hong Kong-Macao Greater Bay Area (Guangdong), Shenzhen 518045. China}

\author{Zidong Lin}
\affiliation{Shenzhen SpinQ Technology Co., Ltd., 518043, Shenzhen, China}

\author{Zheng An}
\affiliation{Department of Physics, The Hong Kong University of Science and Technology, Clear Water Bay, Kowloon, Hong Kong, China}

\author{Guanru Feng}
\affiliation{Shenzhen SpinQ Technology Co., Ltd., 518043, Shenzhen, China}

\author{Zipeng Wu}
\email{z2wu@spinq.cn}
\affiliation{Shenzhen SpinQ Technology Co., Ltd., 518043, Shenzhen, China}

\author{Shiyao Hou}
\email{hshiyao@sicnu.edu.cn}
\affiliation{College of Physics and Electronic Engineering, Center for Computational Sciences, Sichuan Normal University, Chengdu 610068, China}

\author{Jingen Xiang}
\email{jxiang@spinq.cn}
\affiliation{Shenzhen SpinQ Technology Co., Ltd., 518043, Shenzhen, China}


\date{\today}

\begin{abstract}
	Scaling up quantum devices is a central challenge for realizing practical quantum computation. Modular quantum architectures promise scalability, yet experiments to date have relied on either $\lesssim\!10^{3}$-qubit monolithic chips or fragile interconnects with high loss. Here, we introduce a classical linkage scheme that merges multiple independent quantum processing units (QPUs) into a single logical device, enabling thread-level parallelism (TLP).
	Theoretically, we show that quantum routines with product-state inputs and low-rank entangling layers can be re-expressed in an efficient parallelizable form. Experimentally, we validate this architecture on clusters comprising up to sixteen benchtop nuclear magnetic resonance (NMR) quantum nodes. A four-qubit Greenberger-Horne-Zeilinger (GHZ) state is partitioned into parallel two-qubit subcircuits, achieving a fidelity of $93.8\,\%$ with respect to the ideal state. A non-Hermitian evolution, implemented via a truncated Cauchy integral on Hermitian Hamiltonians, reproduces exact observables with high accuracy.
	Our results demonstrate that classical links suffice to scale up the logical size of quantum computations and realize general, non-unitary channels on today's hardware, opening an experimentally accessible route toward software-defined, clustered quantum accelerators.
\end{abstract}


\maketitle

\begin{figure}[!ht]
	\centering
	\includegraphics[width=1\linewidth]{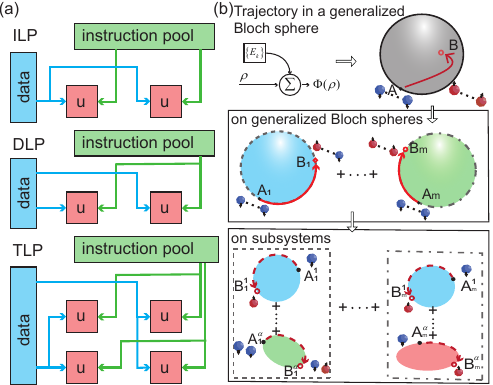}
	\caption{(a) Instruction-, data-, and thread-level parallelism in a classical processor. (b) Factorized evaluation on modular quantum hardware: Trajectory in a generalized Bloch sphere is decomposed into blocks that factorize across subsystems. Local trajectories (on the generalized Bloch sphere of subsystem) are measured independently, and then classically aggregated to recover the global result.}
	\label{fig:Fig1}
\end{figure}

\paragraph*{Introduction.}
For half a century, advances in classical computing have depended critically on increasing parallelism. When single-core performance reached physical and architectural limits, research shifted toward extracting parallelism at multiple levels.
Modern architectures use three typical forms of parallelism
	[see Fig.~\ref{fig:Fig1}(a)]: instruction-level parallelism (ILP) for optimizing execution pipelines, data-level parallelism (DLP) for processing multiple data elements simultaneously, and thread-level parallelism (TLP) for concurrent execution of multiple threads, each handling different data or tasks.
Central processing units (CPUs) exploit ILP through pipelining and out-of-order execution, whereas graphics processing units (GPUs), originally designed for rendering, enable DLP and TLP by integrating thousands of lightweight cores onto a single chip~\cite{Hennessy2011Computer, Kirk2010Programming, Nickolls2010The}. This evolution establishes software-hardware co-design, whereby compute-intensive workloads are automatically partitioned and distributed.

Quantum computing now approaches a similar inflection point. Although the accessible state space of an $n$-qubit device scales as $2^n$, the qubit count of a fabricated chip is fixed; expanding it requires entirely new fabrication procedures.
Increasing qubits sharply elevates risk from materials non-uniformity, wiring and cryogenic I/O scaling, packaging challenges, and yield degradation, rendering large monolithic QPUs economically and technologically daunting.
An alternative is to interconnect existing QPUs, forming clusters that collectively address larger Hilbert spaces and support broader algorithms without redesigning the quantum hardware.
Yet, unlike classical workloads, quantum computation must preserve coherence and entanglement across partitions. Naive partitioning disrupts quantum correlations and is therefore nontrivial.
Recent proposals in distributed quantum computing and quantum networking use long-range entanglement to interconnect distant QPUs~\cite{caleffi2024distributed, monroe2014large, wehner2018quantum}. Despite rapid experimental progress~\cite{li2024heterogeneous, main2025distributed, mollenhauer2025high, liu2024nonlocal}, maintaining reliable quantum links remains challenging.
An alternative is thus to coordinate multiple QPUs using only classical communication, for example, with circuit cutting techniques~\cite{Piveteau2024circuit, Harrow2025optimal, Elisa2024efficient}.
Notably, a 142-qubit superconducting demonstration linked two 127-qubit chips via classical interconnects~\cite{Carrera2024Combining}, and software stacks such as CUDA-Q enable programming across multiple QPUs~\cite{cudaq2024}.
However, existing approaches do not fully exploit parallelism between distinct QPUs, and practical, hardware-level architectures with demonstrations remain scarce [see Appendix~\ref{app:multi_qpu}].

Here, we propose a framework that brings TLP to QPUs. Conceptually, we develop a modular, parallelizable implementation of completely positive trace-preserving (CPTP) channels together with a decomposition that distributes a channel across multiple QPUs with bounded classical coordination overhead.
Architecturally, we realize a scalable cluster that links QPUs using only classical communication and instantiate it with nuclear magnetic resonance (NMR) quantum nodes.
Experimentally, we demonstrate two hallmarks of TLP.
First, we prepare a four-qubit GHZ state using three-qubit NMR nodes, showing that TLP enables emulation of logical systems larger than any individual device. Second, we realize non-Hermitian Hamiltonian simulations, demonstrating the ability to reproduce classes of operations otherwise inaccessible to unitary devices of comparable size. Together, these results indicate that TLP can substantially extend the algorithmic reach of current quantum hardware and offers a concrete route toward scalable, practical quantum computation.

\paragraph*{Architecture.}
In quantum computing, a common task is to compute the expectation value of an observable after applying a quantum channel to an input state. This is specified as
\begin{equation}\label{eq:process}
	\Tr \left(O \cdot \Phi(\rho)\right),
\end{equation}
where $\Phi$ is a quantum channel, $\rho$ is the input state, and $O$ is the observable of interest.
Note that $\rho$ and $O$ play symmetric roles in Eq.~\eqref{eq:process}, enabling parallelism in analogy to classical computing: ILP, by parallelizing operations within $\Phi$; DLP, by running the same operations on multiple copies of $\rho$ or measuring different $O$ in parallel; TLP, by executing multiple quantum operations concurrently, each with different inputs and measurements.
Here we focus on TLP and map a quantum process onto a cluster of modular QPUs.

Frequently, $\rho$ and $O$ factorize across subsystems, i.e., $\rho = \bigotimes_{a=1}^{\mathcal{A}} \rho^{a}$ and $O = \bigotimes_{a=1}^{\mathcal{A}} O^{a}$.
Consider a general form,
\(
\Phi(\cdot)=\sum_{p=1}^{q}
(\sum_{i=1}^{m} c_{p,i}\,U_{p,i})
(\cdot )
(\sum_{j=1}^{m} c_{p,j}\,U_{p,j})^{\!\dagger}
\),
which subsumes standard Kraus and linear combination of unitaries decompositions, defines a CPTP map under appropriate coefficient constraints, and thereby naturally exposes ILP.
Moreover, assume each $U_{p,k}$ admits the factorized expansion,
\begin{equation} \label{U_appro}
	U_{p,k} = \sum_{\alpha=1}^{\ell} \bigotimes_{a=1}^{\mathcal{A}} U^{a}_{p,k\alpha},
\end{equation}
the evaluation thus decouples across subsystems. In particular, Eq.~\eqref{eq:process} can be rewritten as
\begin{equation} \label{eq:parall}
	\sum_{p=1}^q \sum_{i,j=1}^{m} c_{p,i} c_{p,j}^* \sum_{\alpha, \alpha'=1}^{\ell} \prod_{a=1}^{\mathcal{A}} \Tr\left(U^{a}_{p,i\alpha} \rho^{a} U^{a\,\dagger}_{p,j\alpha'} O^{a}\right),
\end{equation}
reducing the computation to a sum of overlaps, each measurable on a small subsystem [see Fig.~\ref{fig:Fig1}(b)]. The final result is obtained by classically aggregating all outputs from the $q \times m^{2} \times \ell^{2} \times \mathcal{A}$ evaluations [see Appendix~\ref{app:detail_arch} for details].
This decomposition provides the theoretical basis for our architecture and experiments. With varying $O^{a}$, it yields a systematic, scalable route to TLP on modular QPU clusters, where heterogeneous operations and measurements are executed in parallel and orchestrated by a classical controller.

As for requirements, beyond tensor-separable $\rho$ and $O$, the derivation of Eq.~\eqref{eq:parall} assumes a factorization of the global unitary into sums of local operators, as in Eq.~\eqref{U_appro}. A brute-force operator-Schmidt decomposition is possible in principle, but the cost scales as $O(d^{2n})$ for $n$ qubits and quickly becomes impractical. Fortunately, layered circuit structures are common in practice. If $U_{p,k}$ admits a layered-circuit form, the factorization can be obtained efficiently and even optimized by recasting the task as a min-cut problem on the associated circuit graph~\cite{Arora2009Expander, Monika2024Deterministic}. Consider the bisection min-cut problem, i.e., $\mathcal{A}=2$.
A layered circuit $U_{p,k}$ can be written as
$ \prod_{t=1}^{m_{d}} \!(\sum_{\alpha_t=1}^{\ell_t}\bigotimes_{a=1}^{2} g_{a}^{\alpha_t} U^{(t)}_{a})$, where the $t$-th entangling layer factorizes as
$  \sum_{\alpha_t=1}^{\ell_t}
	\bigotimes_{a=1}^{2} g^{a}_{t\alpha_t}$.
In fact, some $g^{a}_{t\alpha_t}$ may be the identity. In this case, $m_{d}$, originally set as the number of circuit layers, can be reduced.
The goal is to bipartition the circuit while minimizing the number of inter-partition entangling gates.
Applying a circuit-graph min-cut algorithm [complexity $O(n^{2})$ in the number $n$ of graph vertices, specified in Appendix~\ref{app:assumption}.], one finds a cut of size $m' \le m_{d}$ such that the global expectation in Eq.~\eqref{eq:parall} factorizes into at most
$q \times m^{2} \times
	({\textstyle\prod_{t=1}^{m'}}\ell_t)^{2}
	\times 2$
independent single-subsystem traces, each executable on an individual QPU node. Furthermore, each trace of overlap in Eq.~\eqref{eq:parall} can be measured efficiently using a single-ancilla protocol, requiring only one additional qubit per subsystem; see Appendix~\ref{app:assumption}, especially Lemmas~\ref{prop:layered-decomp} and~\ref{prop:single-ancilla}, for details.

\begin{figure}[!htbp]
	\centering
	\includegraphics[width=1\linewidth]{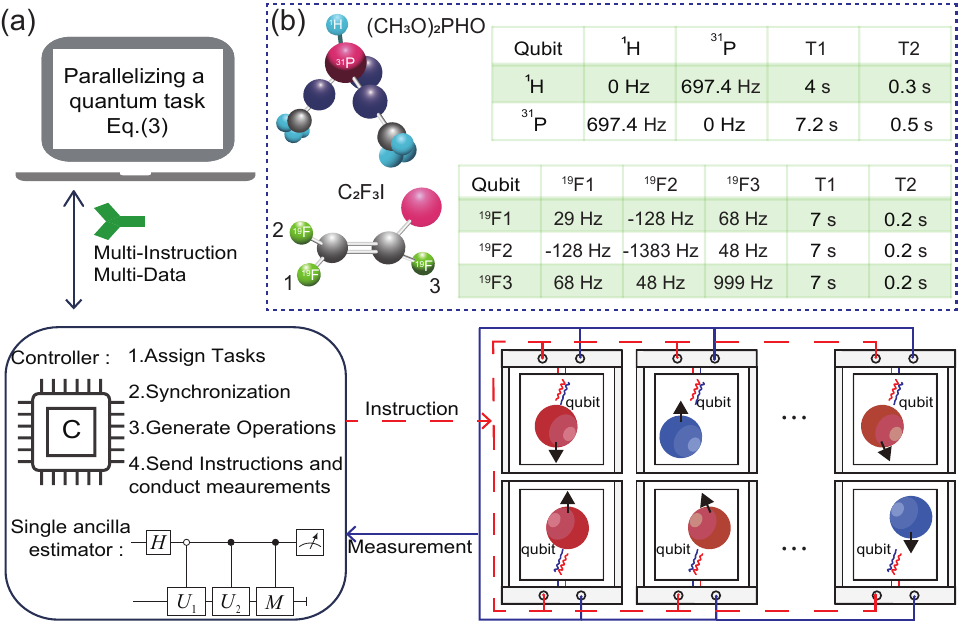}
	\caption{Hardware overview. (a) Clustered-QPU architecture enabling TLP in quantum computing. (b) Physical sample assembly employed in this work.}
	\label{fig:Fig2}
\end{figure}

\paragraph*{Hardware.} Fig.~\ref{fig:Fig2}(a) shows a cluster with loose coupling, where multiple QPUs are integrated. At the core of the system is a classical controller that assigns computational tasks, synchronizes all QPU nodes, generates pulse operations, and coordinates instruction dispatch and measurement collection. Operating in a multi-instruction, multi-data mode, the controller ensures both flexibility and high throughput for realizing TLP on quantum devices.
Each QPU node can, in principle, be instantiated on any platform, such as nuclear-spin qubits, photonic qubits, superconducting circuits, or others, provided it supports universal control and high-fidelity state preparation and measurement. This modular design enables seamless integration of heterogeneous quantum processors within a unified computational platform, with all inter-processor signaling carried classically.

During operation, the central scheduler decomposes tasks according to the theoretical framework, dispatches the resulting subtasks to each node, and orchestrates control-pulse timing and data acquisition to maintain precise synchronization. The cluster is fully modular and plug-and-play: nodes can be added, removed, or replaced on the fly, with the controller automatically updating task assignments and preserving system-wide logical consistency. Because all inter-node communication is classical, the architecture is inherently robust and highly scalable.

As a demonstration, we construct clusters comprising 8 or 16 benchtop NMR QPUs~\cite{hou2021spinq, Feng2022SpinQ}, where each node hosts 3 or 2 nuclear-spin qubits [Fig.~\ref{fig:Fig2}(b)].
The physical realization employs different nuclear species, $^{1}$H, $^{19}$F, and $^{31}$P, each with distinct Larmor frequencies, chemical shifts, and coherence properties. Key parameters, including $T_1$ and $T_2$, are summarized in Fig.~\ref{fig:Fig2}(b) and Appendix~\ref{app:hardware}, highlighting multi-second coherence that supports reliable thread-level parallel computation.
This platform validates our architecture along two axes. First, clusters of small QPUs collectively emulate circuits that exceed the size of any single node. Second, general quantum channels are realized efficiently by parallelizing simple unitary evolutions. We next present two representative experimental tasks.

\begin{figure}[!htbp]
	\centering
	\includegraphics[width=1\linewidth]{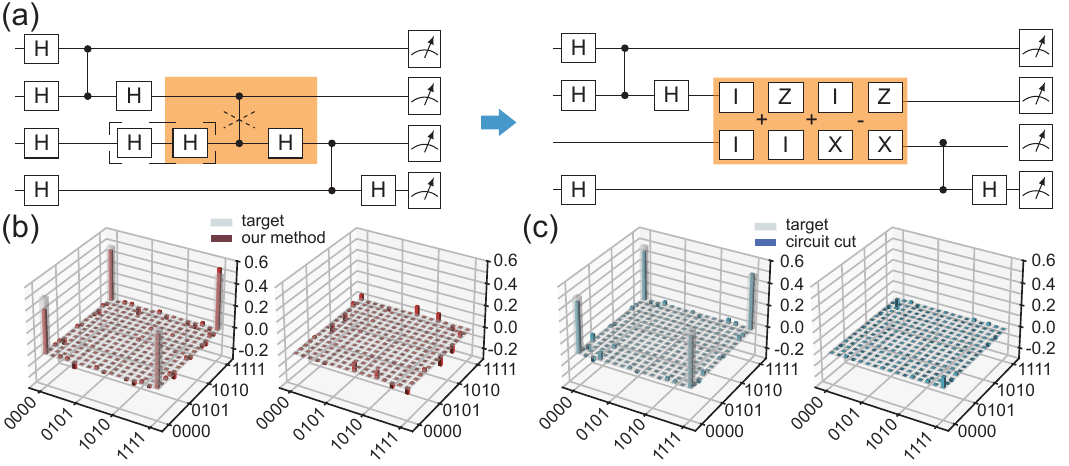}
	\caption{(a) Experimental quantum circuit for preparing a four-qubit GHZ state. The CZ gate between qubits 2 and 3 has been ``cut'' and implemented as a linear combination of local operations. (b) and (c) depict real and imaginary parts of the reconstructed density matrices for our method and the circuit-cut state, with the target value is shown with transparent blocks.}
	\label{fig:Fig3}
\end{figure}

\paragraph*{GHZ state preparation.}
We first demonstrate the preparation of a four-qubit GHZ state, a canonical benchmark in quantum information~\cite{pont2024high, bao2024creating}. As shown in Fig.~\ref{fig:Fig3}(a), the CNOT between qubits 2 and 3 is ``cut'' and replaced by a linear combination of local operations,
$U_{cnot} = \frac{1}{2} \left(\openone \otimes \openone + \sigma_z \otimes \openone + \openone \otimes \sigma_x - \sigma_z \otimes \sigma_x\right),$
where $\sigma_{x,z}$ denote Pauli operators and $\openone$ is the identity. This allows the output state to be expressed as a sum of product substates, and the original 4-qubit GHZ circuit is thus decomposed into a superposition of two-qubit subcircuits, making the protocol compatible with small-scale hardware. Using the single-ancilla estimator, we measure the overlaps and implement parallelization on two kinds of circuits, shown in Appendix~\ref{app:exp1}.
Additionally, we show compatibility of the hardware with another technique invoking parallelization, quantum circuit cutting~\cite{mitarai2021constructing}, which decomposes the four-qubit circuit into a sum of ten two-qubit subcircuits.

Both approaches are implemented on clusters with either eight 3-qubit nodes or sixteen 2-qubit nodes. To obtain full information on the prepared GHZ state, we perform 128 observable evaluations, extracting $\langle \sigma_x \rangle $, $\langle \sigma_y\rangle$, and the projectors $\ket{0}\bra{0}$, $\ket{1}\bra{1}$ on the ancilla qubit across 32 circuits derived from two template circuits. In the circuit-cutting scheme, full state tomography is employed, yielding 160 observable evaluations from 10 circuits. The reconstructed density matrices are shown in Fig.~\ref{fig:Fig3}(b,c), where techniques~\cite{Huggins2021Virtual} are employed to mitigate incoherent errors as quite different effects caused by homo ((CH$_3$O)$_2$PHO) and hetero (C$_2$F$_3$I) nuclear systems, and fidelities are calculated as 93.8\% and 92.3\% relative to the ideal GHZ state, respectively. By executing heterogeneous operations and measurements in parallel, these results validate TLP on our clusters and enable the preparation of target states beyond the qubit capacity of any single node. Full experimental details are provided in Appendix~\ref{app:exp1}.

For comparison, circuit cutting aggregates subresults classically and therefore requires no ancilla to probe overlaps. However, it entails more subtasks (160 vs.~128 in our experiment) and heavier preprocessing to decompose a general entangling gate. Under the condition that the entangling gate is fixed as CNOT, as the number $m$ of layers or inter-partition entangling gates increases, the subtask counts scale as $16\times10^m \mbox{~vs.~} 8\times2^m$, underscoring the efficiency advantage of our approach to TLP. Moreover, our architecture, with single-ancilla estimator, benefits the implementation of general quantum channels, as demonstrated below.

\paragraph*{Non-Hermitian system simulation.}
Efficient Hamiltonian simulation on quantum devices promises deep insights into quantum dynamics with broad applications~\cite{wang2021simulating,bosse2025efficient}. The GHZ preparation above illustrates a lightweight use of our multi-QPU cluster (of the order of $10^2$ subtasks). Here we present a more demanding application involving the order of $10^3$ subtasks executed across the $16\times2$-qubit cluster. Focusing on non-Hermitian Hamiltonians and imaginary-time evolution, we employ TLP across QPUs, implementing the linear-combination Hamiltonian simulation (LCHS) method~\cite{an2023linear}.

\begin{figure}[!htbp]
	\centering
	\includegraphics[width=1\linewidth]{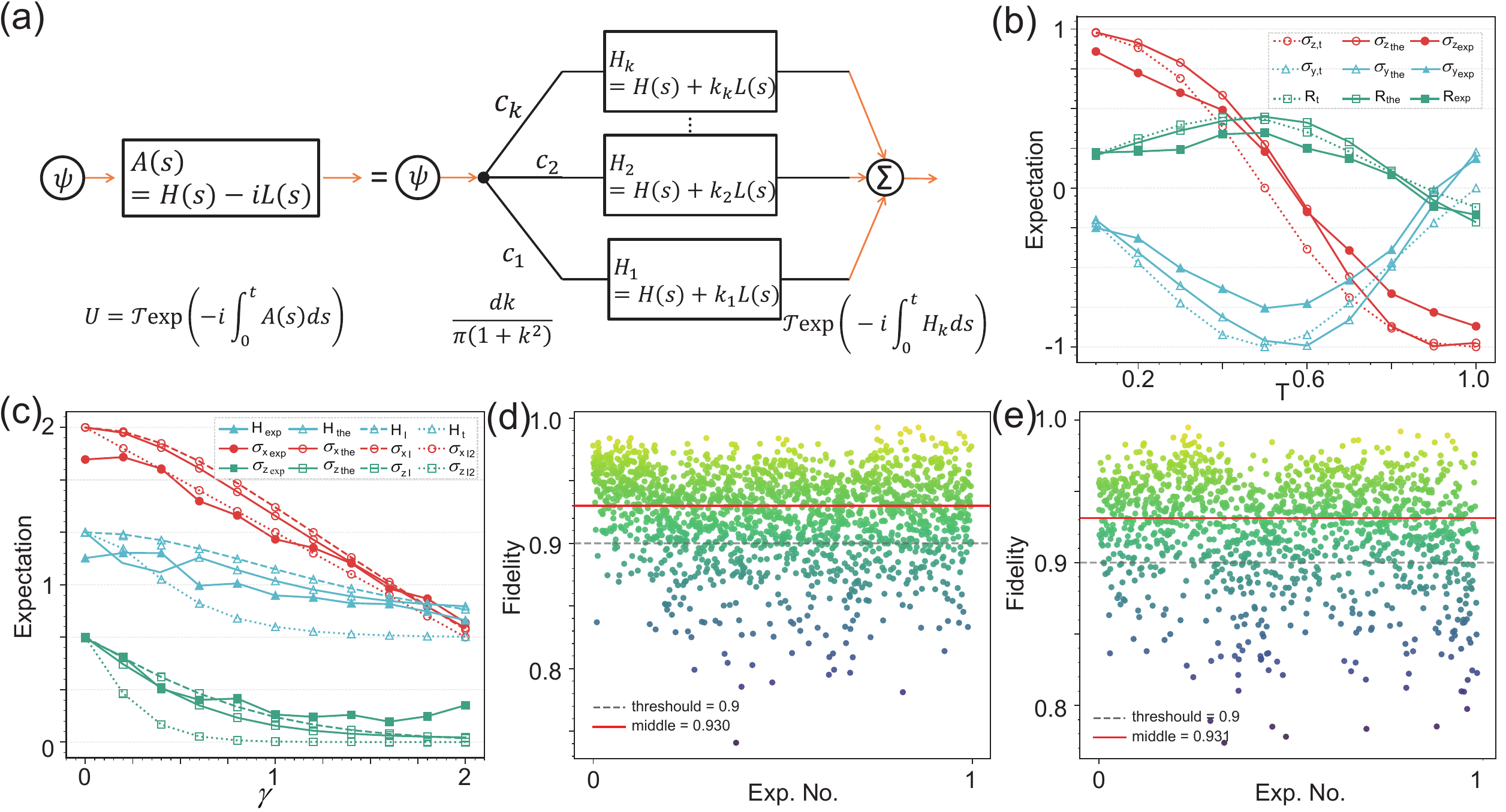}
	\caption{(a) sketches the basic idea of linear combination of Hamiltonian simulations. (b) Time evolution of $\langle\sigma_{y,z}\rangle$ and a randomly generated Hermitian observable under a non-Hermitian Hamiltonian. (c) Measured $\langle H(\gamma)\rangle$, and $\langle\sigma_{x,z}\rangle$ after imaginary-time evolution of $H(\gamma)$. Solid lines with filled points indicate experimental data; hollow points with solid represents simulation result via experimental method; while dashed lines represent results via first principles calculations. In (c) dashed-dotted lines represent results from exact diagnolazation and longer-time imaginary-time evolution. (d) and (e) show the fidelity of the experimentally prepared states relative to the numerical predictions, while the median fidelity and the 90 \% threshold are highlighted.}
	\label{fig:Fig4}
\end{figure}

Consider $A(t)=H(t)-iL(t)$ with the Hermitian components $H(t)$ and $L(t)$. LCHS is built on the identity, shown in Fig.~\ref{fig:Fig4}(a),
\(
\mathcal{T}\exp\!\left(- i\int_0^t\! A(s)\,ds\right)
=\!\int_{\mathbb{R}}\!\frac{dk}{\pi(1{+}k^2)} \mathcal{T}\exp\!\left(\!-i\int_0^t\![H(s){+}kL(s)]\,ds\right),
\)
where $\mathcal{T}$ denotes time ordering and the Cauchy--Lorentz kernel $1/\!(\pi(1{+}k^2))$ arises from the Fourier representation of $e^{-|x|}$.
Discretizing the $k$-integral on $[-K,K]$ with uniform step $\Delta k$, quadrature nodes $\{k_j\}$, and trapezoidal weights $\{w_j\}$ yields
\begin{equation}
	u(t)\approx\sum_j c_j U_j(t),
	\label{eq:lchs-disc}
\end{equation}
with $U_j(t)=\mathcal{T}\exp(-i\int_0^t [H(s){+}k_jL(s)]ds)$ and
$c_j=w_j/(\pi(1{+}k_j^2))$.
Hence the quantum operation governed by $A(t)$ fits as a special case of Eq.~\eqref{eq:parall}.
If only expectation values are needed,
\(\langle O\rangle_t=\langle u(t)|O|u(t)\rangle\approx
\sum_{k,k'} c_k^{*}c_{k'} \langle u_0|\,U_k^\dagger(t)\,O\,U_{k'}\ket{u_0}\),
which is a subset of Eq.~\eqref{eq:parall}, with $\ket{u_0}$ the initial state and $|u(t)\rangle$ the time-evolved state. In our TLP setting, these terms are dispatched across QPU threads and aggregated classically.

For the experiment, we first consider dynamics generated by a single-qubit and reproduce time-independent non-Hermitian Hamiltonian simulation with $H = \sigma_x$ and $L = \openone + \sigma_z$~\cite{bender2003must, wu2019observation}.
Here $\sigma_{x,z}$ are Pauli matrices and $\openone$ is the identity. Initializing the system in $\ket{0}$, we record $\langle \sigma_{x,z} \rangle$ as functions of the evolution time.

Based on LCHS, the non-Hermitian Hamiltonian dynamics decomposes into a collection of parallel subtasks (approximately $400T^2$, with $T$ the evolution time) implemented using a single-ancilla estimator circuit. Each subtask corresponds to one experiment and is executed on a node of the $16\times2$-qubit NMR-QPU cluster. In total, we perform over 1700 experimental runs (about 106 per node) to resolve the time evolution($T=0.1\,k (k=1,\dots,10)$), thereby demonstrating TLP. For benchmarking, we also carry out numerical simulations using both the LCHS method and direct integration of the Schr{\"o}dinger's equation. The two approaches agree with fidelity 99.5\%, corroborating the accuracy of LCHS.
Fig.~\ref{fig:Fig4}(b) shows the time evolution of $\langle \sigma_{y,z} \rangle$, obtained from LCHS experiments, LCHS simulations, and direct Schr{\"o}dinger integration. The absolute deviation is $0.129\pm0.070$. The close agreement across all traces demonstrates the feasibility and reliability of parallel non-Hermitian Hamiltonian simulation on our QPU cluster. Further evidence is provided in Fig.~\ref{fig:Fig4}(d), which reports the state fidelity of each experimental snapshot relative to theory: the median fidelity is 93.1\%, and a 90\% threshold is indicated. See Appendix~\ref{app:exp2} for details.

Then we demonstrate a ground-energy estimation via imaginary-time evolution for $H(\gamma)=2\openone+\gamma\sigma_x, \gamma\in[0,2]$, whose analytic ground energy is $E_0(\gamma)=2-|\gamma|$.
Starting from $\ket{0}$, we apply non-unitary evolution $e^{-H(\gamma)T}$ followed by renormalization, which approximates the ground state for sufficiently large $T$. The ground energy is then estimated by the expectation $\langle H(\gamma)\rangle$. This setting provides a clean testbed for imaginary-time simulation, given the closed-form $E_0(\gamma)$ for benchmarking.

Similar to LCHS and more precisely via a truncated Cauchy integral representation~\cite{Huo2023errorresilientmonte}, we approximate the imaginary-time propagator as a collection of parallel subtasks ($\sim 121$ per $\gamma$). We demonstrate TLP by executing over 1200 subtasks with a single-ancilla estimator circuit on the $16\times2$-qubit NMR cluster. In experiments, we let $\gamma = 0.2\,k$ ($k=0,\dots,10$), and measure $\langle H(\gamma)\rangle$, $\langle \sigma_{x, y}\rangle$ with $T=0.5$.
For benchmarking, experimental results are compared against three references: a LCHS simulation, direct imaginary-time evolution with $T=0.5$ and exact diagonalization of $H(\gamma)$ (and, as a tighter baseline, imaginary-time evolution with $T=1.5$).
Fig.~\ref{fig:Fig4}(c) shows the measured and simulated expectation values of $\langle \sigma_{x,z} \rangle$, and the ground energy as a function of $\gamma$.
Experimental results, simulations of the experimental method, numerical imaginary-time evolution, and exact diagonalization or long-time (e.g., $T=1.5$) imaginary-time evolution are shown with the absolute deviation $0.136\pm 0.088$, which confirms that our parallel QPU cluster can efficiently and accurately realize imaginary-time evolution and ground-state energy estimation. Additional support is given by Fig.~\ref{fig:Fig4}(e), which summarizes the fidelity between the experimentally prepared states and theoretical ground states for each $\gamma$. With a moderate total evolution time $T=0.5$, all data points achieve fidelities exceeding 99\%. Increasing $T$ (e.g., $T=1.5$) further improves the ground-state approximation but requires more experimental runs, reflecting the tradeoff between accuracy and experimental cost [See Appendix~\ref{app:exp2} for details].

\paragraph*{Conclusion.}
While building large, general-purpose QPUs remains technologically and economically challenging, clustering existing devices enables significant performance enhancement without waiting for next-generation hardware. Even as large, fault-tolerant chips become available, their re-fabrication will likely remain costly, making QPU clustering a viable ``performance boost" strategy, analogous to overclocking in classical systems, but without imposing extra physical strain on individual chips.

Our approach harnesses the strengths of current quantum processors while respecting their limitations, enabling flexible parallel execution of diverse quantum operations and measurements. We show that quantum routines with product-state inputs and low-rank entangling layers can be re-expressed in an efficient parallelizable form. By orchestrating many small QPUs in parallel, we achieve TLP and scalability unattainable by a single device. This architecture was validated experimentally using NMR clusters of eight and sixteen nodes. Lightweight benchmarks distributed a four-qubit GHZ circuit across two-qubit nodes, reproducing the target state with $93.8\%$ fidelity. To further stress-test the system, we implemented a resource-intensive protocol involving thousands of parallel executions for LCHS, successfully emulating both non-Hermitian and imaginary-time dynamics-constituting the first such demonstration on a modular quantum platform.

In summary, TLP on quantum devices immediately enlarges the logical register and enables complex channel-level operations inaccessible to standalone chips. This framework thus benefits other near-term quantum applications, such as variational quantum eigensolvers~\cite{Mineh2023Accelerating, Cattelan2025Parallel}. Since our scheme relies only on classical interconnects, it is compatible with current fabrication capabilities and can be scaled, offering a pragmatic and flexible bridge from today's few-qubit devices to future large-scale, fault-tolerant quantum accelerators.

\section*{acknowledgments}
We thank texra.ai for helpful suggestions on writing and language.
K.L. and S.H. acknowledge the Scientific Foundation for Youth Scholars of Shenzhen University, Guangdong Provincial Quantum Science Strategic Initiative (GDZX2403001, GDZX2303001).

\bibliographystyle{apsrev4-2}
\bibliography{apssamp} 
 
\onecolumngrid
\appendix
\newpage
\setcounter{figure}{0}
\renewcommand\thefigure{S\arabic{figure}}
\setcounter{table}{0}
\renewcommand\thetable{S\arabic{table}}

\section{Supplementary information for multi-QPU strategies}
\label{app:multi_qpu}
When several quantum-processing units (QPUs) are asked to execute a
single workload, two fundamentally different resources can be used to
tie them together:
\begin{enumerate}
	\item \textit{Quantum correlated links ---} Entanglement is distributed between modules and exploited through gate teleportation, remote measurement, or lattice surgery. Theoretical blueprints date back to modular ion-trap architectures with photonic routers~\cite{monroe2014large} and underpin the long-term vision of distributed quantum computing~\cite{li2024heterogeneous, main2025distributed, mollenhauer2025high} or a quantum internet~\cite{wehner2018quantum,liu2024nonlocal}.
	\item \textit{Classically correlated links ---}  Each QPU is measured locally; the global observable is then reconstructed from the measurement records with ``divide-and-conquer'' methods. This family includes wire- and gate-based quantum circuit cutting~\cite{Harrow2025optimal}, circuit knitting~\cite{Piveteau2024circuit} and dynamic circuits techniques\cite{Elisa2024efficient, Carrera2024Combining}.
\end{enumerate}

Both routes trade resources differently. Quantum links preserve coherence across modules and avoid exponential sampling overhead, but they impose stringent latency and loss budgets on the interconnect. Classical links need no fragile quantum channel and can already incorporate non-unitary dynamics, yet their classical post-processing overhead grows with every wire or gate that is cut.
Specifically, Table~\ref{tab:multi_qpu_summary} briefly summarizes these representative multi-QPU strategies and their key features.
\begin{table}[ht]
	\centering
	\caption{\textbf{Representative multi-QPU strategies and their key features}}
	\label{tab:multi_qpu_summary}
	\footnotesize
	\setlength{\tabcolsep}{4pt}
	\begin{adjustbox}{max width=0.95\linewidth,center}
		\begin{tabularx}{1.2\linewidth}{@{}lcccccc@{}}
			\toprule
			\textbf{Work}                                       & \textbf{Year}           & \textbf{Correlation}               &
			\textbf{Interconnect}                               & \textbf{Non-unitary}    &
			\textbf{Key feature}                                & \textbf{Hardware scale}                                                                                                                                   \\
			\midrule
			\multicolumn{2}{l}{Distributed Qauntum Computing}   &                         &                                    &                               &                                                            \\
			Monroe \textit{et al.}~\cite{monroe2014large}       & 2014                    & Quantum                            &Probabilistic photonic interface                   & n/a & Modular blueprint            & Proposal              \\
			Spin-photon CMOS~\cite{li2024heterogeneous}         & 2024                    & Quantum                            & photonics                     & No  & Wafer-scale spin-photon node & 1 mm$^{2}$ chip       \\
			Trapped-ion link~\cite{main2025distributed}         & 2025                    & Quantum                            &
			Heralded photons                                    & No                      & Deterministic gate teleport        & $2$ ions module (\(\sim\) 2m)                                                                         \\
			Microwave cable~\cite{mollenhauer2025high}          & 2025                    & Quantum                            &
			Plug-and-play coaxial cable                                  & No                      & 99 \% swap in 100ns        & $2$ transmons(\(\le\) 1m)                                                                     \\
			\hline
			\multicolumn{2}{l}{Quantum Internet}                &                         &                                    &                               &                                                            \\
			Optical 7 km gate~\cite{liu2024nonlocal}            & 2024                    & Quantum                            &
			Telecom fiber                                       & No                      & Metro-scale entangling gate        & 2 memories                                                                                 \\
			\hline
			\multicolumn{2}{l}{Dynamic circuits}                &                         &                                    &                               &                                                            \\
			Dynamic circuits~\cite{Elisa2024efficient}          & 2024                    & Classical                          &
			Conditional feed-forward operations                                          & No                      & 101-qubit CNOT teleport            & 127-qubit chip                                                                             \\
			Dynamic-circuit cutting~\cite{Carrera2024Combining} & 2024                    & Classical                          & Real-time classical link                   & No  & 142-qubit graph via cutting  & $2{\times}127$ qubits \\
			\hline
			\multicolumn{2}{l}{Quantum circuit cutting}         &                         &                                    &                               &                                                            \\
			Optimal circuit cuts~\cite{Harrow2025optimal}       & 2025                    & Classical                          & Theory                        & No  & Tight sample-cost bounds     & Theoretical           \\
			\hline
			\multicolumn{2}{l}{Quantum circuit knitting}        &                         &                                    &                               &                                                            \\
			Circuit knitting~\cite{Piveteau2024circuit}         & 2024                    & Classical                          &
			Software only                                       & No                      & Reducing gates overhead                   & Simulator                                                                                  \\
			\hline
			\multicolumn{2}{l}{Others}                          &                         &                                    &                               &                                                            \\
			CUDA-Q multi-QPU~\cite{cudaq2024}                   & 2024                    & Classical                          &
			Software only                                       & n/a                     & Async kernels; More tests required & Simulator                                                                                  \\
			\textbf{This work}                                  & 2025                    & Classical                          &
			Ethernet                                            & \textbf{Yes}            & Efficient in a large class of CPTP map.           & $16{\times}2/8\times 3$ spins                                                              \\
			\bottomrule
		\end{tabularx}
	\end{adjustbox}

\end{table}

\paragraph{Quantum-correlated prototypes.}
\begin{itemize}
	\item Spin-photon CMOS integration~\cite{li2024heterogeneous} - Li \textit{et al.} heterogeneously bond diamond colour-centre qubits onto 45 nm CMOS photonics, enabling wafer-scale quantum light sources and detectors.
	\item Trapped-ion optical link~\cite{main2025distributed} - Two ion chains 2 m apart execute a distributed Grover search via deterministic photonic gate teleportation, reaching 86 \% gate fidelity.
	\item  Plug-and-play microwave cable~\cite{mollenhauer2025high}  - A detachable coaxial bus swaps microwave photons between two superconducting chips in $<\!100$ ns with $>99\%$ efficiency.
	\item  7\,km non-local photonic gate~\cite{liu2024nonlocal}  - Quantum memories plus telecom photons realise entangling gates over a 7 km fibre loop.
\end{itemize}

\paragraph{Classically correlated prototypes.}
\begin{itemize}
	\item Dynamic-circuit cutting - Shallow measure-and-feed-forward protocols teleport CNOTs across 100+ qubits \cite{Elisa2024efficient}, while two 127-qubit processors linked only by real-time classical feed-forward prepare 142-qubit graph states~\cite{Carrera2024Combining}.
	\item Circuit knitting with LOCC~\cite{Piveteau2024circuit} - Information-theoretic analysis shows classical messaging reduces sampling overhead for knitted circuits.
	\item Optimal quantum circuit cuts~\cite{Harrow2025optimal} - Tight bounds on sample cost and an application to clustered Hamiltonian simulation.
\end{itemize}

\paragraph{This work - thread-level parallelism for quantum channels.}%
We introduce a method that factorizes an arbitrary completely-positive trace-preserving map into independent Kraus blocks, dispatches those blocks to a cluster of QPUs over standard Ethernet, and recombines the measurement records in $\mathcal{O}(1)$ classical time. Unlike prior ``divide-and-conquer'' methods that target only unitary circuits~\cite{Carrera2024Combining,Elisa2024efficient}, our scheme natively supports non-unitary dynamics such as non-hermitian Hamiltonian evolution and imaginary-time propagation.

\textit{Hardware demonstration---}  A $16\times2$-qubit (or $8\times3$-qubit) NMR cluster prepares a 4-qubit GHZ state with a tomographic fidelity of $93.8\,\%$; emulates non-Hermitian Hamiltonian and imaginary-time dynamics.
These results establish that purely classical networking can extend both the size and the algorithmic richness of NISQ workloads beyond what any single device can achieve today, delivering the first channel-level TLP benchmark on real hardware.

\section{Supplementary information for realizing thread level parallelism}
\label{app:detail_arch}
We start from the generic quantum-computing objective
\[
	\Tr\![ O \Phi(\rho) ],
\]
where $\Phi$ is a quantum channel, $\rho$ an input state and $O$ the observable of interest.

$\Phi$ can be a general structure that admits a linear-combination-of-unitaries form
\begin{equation}\label{eq:LCU}
	\Phi(\,\cdot\,) =
	\sum_{p=1}^{q}\left(\sum_{i=1}^{m} c_{p,i}\,U_{p,i}\right)
	(\,\cdot\,)
	\left(\sum_{j=1}^{m} c_{p,j}\,U_{p,j}\right)^{\!\dagger},
\end{equation}
with complex amplitudes $c_{p,i}$ and unitaries $\{U_{p,i}\}_{i=1}^{m}$. Substituting Eq.~\eqref{eq:LCU} into the trace and expanding gives
\begin{align}
	\Tr\![ O\,\Phi(\rho) ]
	 & =\sum_{p=1}^{q}
	\sum_{i,j=1}^{m}
	c_{p,i}\,c_{p,j}^{*}
	\Tr\!\bigl[
	O U_{p,i} \rho\,U_{p,j}^{\dagger}
	\bigr].
	\label{eq:double-sum}
\end{align}
Here, we employ a without losing generosity assumption that both the state and the observable are product states over $\mathcal{A}$ identical subsystems,
\[
	\rho =  \bigotimes_{a=1}^{\mathcal{A}}\rho^{a},
	\qquad
	O      =  \bigotimes_{a=1}^{\mathcal{A}}O^{a}.
\]
And a non-trivial requirement that every unitary $U_{p,i}$ have an explicit expression of the summation of tensor-product decomposition,
\begin{equation}\label{eq:U-decomp}
	U_{p,i} =
	\sum_{\alpha=1}^{\ell}
	\bigotimes_{a=1}^{\mathcal{A}} U^{a}_{p,i\alpha},
\end{equation}
with the same $\ell$ for all indices. Insert Eq.~\eqref{eq:U-decomp} (and its Hermitian conjugate) into Eq.~\eqref{eq:double-sum}; linearity yields
\begin{align}
	\Tr\![ O\,\Phi(\rho) ]
	 & =\sum_{p=1}^{q}\sum_{i,j=1}^{m} c_{p,i}\,c_{p,j}^{*}
	\sum_{\alpha,\alpha'=1}^{\ell}
	\Tr\!\Bigl[
		O
		\Bigl(    \bigotimes_{a}U^{a}_{p,i\alpha}
		\Bigr)
		\rho
		\Bigl(    \bigotimes_{a}U^{a}_{p,j\alpha'}
		\Bigr)^{\!\dagger}
		\Bigr].
	\label{eq:expand}
\end{align}
Because both the state, observable and each tensor factor act on disjoint Hilbert spaces, the global trace factorizes into a product of local traces:
\begin{align}
	\Tr\!\Bigl[
		O
		\bigl(    \bigotimes_{a}U^{a}_{p,i\alpha}
		\bigr)
		\rho
		\bigl(    \bigotimes_{a}U^{a}_{p,j\alpha'}
		\bigr)^{\!\dagger}
		\Bigr]
	=
	\prod_{a=1}^{\mathcal{A}}
	\Tr\!\bigl[
	U^{a}_{p,i\alpha} \rho^{a}\,U^{a\,\dagger}_{p,j\alpha'}\,O^{a}
	\bigr].
\end{align}
Substituting this identity into Eq.~\eqref{eq:expand} yields the fully
factorised expression
\begin{equation}\label{eq:parall-derived}
	\sum_{p=1}^{q}\sum_{i,j=1}^{m} c_{p,i}\,c_{p,j}^{*}
	\sum_{\alpha,\alpha'=1}^{\ell}
	\prod_{a=1}^{\mathcal{A}}
	\Tr\!\bigl[
	U^{a}_{p,i\alpha} \rho^{a}\,U^{a\,\dagger}_{p,j\alpha'}
	\,O^{a}
	\bigr],
\end{equation}
which is exactly Eq.~\eqref{eq:parall} in the main text. Each factor inside the product is measurable on a single subsystem with programmable unitary transformations and measurements, while the outer sums can be accumulated classically. This indicates thread-level parallelism, TLP, can be realized across quantum processors.

In summary, every quantum-information task of the form $\Tr\![\,O\,\Phi(\rho)]$ can be split into $q \times m^{2} \times \ell^{2} \times \mathcal{A}$ independent sub-tasks, provided that the state $\rho$ and observable $O$ factorize over $\mathcal{A}$ subsystems, and each  operator \(U_{p,k}\) admits the tensor-product expansion of Eq.~\eqref{eq:U-decomp}.
	The integers $q$, $m$, $\ell$ and $\mathcal{A}$ are precisely those introduced in Eqs.~\eqref{eq:LCU} and~\eqref{eq:U-decomp}. Furthermore, if the channel reduces to an incoherent sum of unitaries, $\Phi(\rho)=\sum_{k=1}^{q} U_{k} \rho\,U_{k}^{\dagger},$ and every $U_{k}$ further decomposes as \( U_{k}= \sum_{\alpha=1}^{\ell} \bigotimes_{a=1}^{\mathcal{A}} U^{a}_{k\alpha},\) then the workload factorizes into only $q \times \ell^{2} \times \mathcal{A}$ sub-tasks, because the double index $(i,j)$ collapses to a single index $k$ ($m=1$).

\section{Supplementary information for pre-conditions in architecture}
\label{app:assumption}
The first is an efficient factorization that rewrites the global unitary into sums of local operators.
A brute-force operator-Schmidt decomposition would accomplish factorization step, its cost grows as $\Theta(d^{2n})$ for $n$ qubits and is therefore intractable.

Instead, we exploit the layered structure of quantum circuits to recursively apply precomputed local factorizations. In general, given a depth-$m_d$ circuit
\[
	U =  \prod_{i=1}^{m_d}
	\Bigl(          G^{(i)}_{\!\mathrm e}
	\otimes_{a=1}^{\mathcal A} U^{(i)}_{a}
	\Bigr),
\]
whose $i$-th layer contains a single entangling block
$G^{(i)}_{\!\mathrm e}$ acting across the $\mathcal A$ subsystems,
and a factorized expansion of each entangling block,
\[
	G^{(i)}_{\!\mathrm e}
	=
	\sum_{\alpha_i=1}^{\ell_i}
	\bigotimes_{a=1}^{\mathcal A} g^{a}_{i\alpha_i},
	\qquad
	g^{a}_{i\alpha_i}\in\mathsf U(\dim\mathcal H_a),
\]
we can rewrite the entire circuit as
\[
	U =
	\prod_{i=1}^{m_d}
	\sum_{\alpha_i=1}^{\ell_i}
	\bigotimes_{a=1}^{\mathcal A}
	g^{a}_{i\alpha_i}\,U^{(i)}_{a},
\]
i.e.\ as a nested linear combination of $\prod_{i=1}^{m_d}\ell_i$ purely local tensor-product operators.

Bisecting the circuit is a good start of the problem.
It is trivial that one have a way to bisect a unitary into sub circuits. For current popular circuit, entangling block can be chosen as CNOT or CZ, which have a fixed factorized expansion.
Therefore, $m_d$ appears , $\prod_{i=1}^{m_d}\ell_i = \ell$, which influences the number of sub tasks take this form. To optimize it, we introduce an algorithm of layer-wise decomposition (Lemma~\ref{prop:layered-decomp}) to reduce it.

\begin{lem}[Layer-wise decomposition]
	\label{prop:layered-decomp}
	Let
	\[
		U
		=\prod_{i=1}^{m_d}
		\Bigl(G^{(i)}_{e} \
		\bigotimes_{a=1}^{n}U^{(i)}_{a}\Bigr)
	\]
	be an $n$-qubit circuit of depth~$m_d$, where at each layer~$i$, $G^{(i)}_{e}$ is a set of two-qubit gates that may couple any pair of qubits, and $U^{(i)}_{a}$ acts locally on qubit $a$.
	\begin{enumerate}[label=(\roman*)]
		\item\label{item:cut}
		      Viewing the circuit as a weighted graph $G=(V,E)$ with $|V|=n$ and edge weights $w_{uv}$ equal to the number of two-qubit gates acting on $\{u,v\}$, one can find in $\tilde O(|E|)\subseteq\tilde O(n^{2})$ time (Henzinger-Li-Rao-Wang~\cite{Monika2024Deterministic}) a minimum cut. The cut removes $m'\le m_d$ crossing gates $\{g^{(i)}\}_{i=1}^{m'}\subseteq\{G^{(i)}_{e}\}$ and partitions the qubits into two subsystems, hereafter indexed by $a\in\{1,2\}$.
		\item\label{item:factor}
		      Suppose each crossing gate admits a known factorization
		      $g^{(i)} =\sum_{\alpha_i=1}^{\ell_i} V^{(1)}_{i\alpha_i}\otimes V^{(2)}_{i\alpha_i},$
		      where all $V^{(a)}_{i\alpha_i}$ are unitary ($\ell_i\le 4$ for \textsc{cnot}/\textsc{cz}). Denoting \(\tilde{U}_{a}^{(i)}\) as the local operation within subsystem $a$, the entire circuit decomposes as
			      \[ U = \prod_{i=1}^{m'} \sum_{\alpha_i=1}^{\ell_i}\bigotimes_{a=1}^{2} V^{(a)}_{i\alpha_i} \tilde{U}_{a}^{(i)}, \]
			      which contains at most $\prod_{i=1}^{m'}\ell_i$ tensor terms, independent  of the dimensions of the two subsystems.
	\end{enumerate}
\end{lem}

Remarkably, if each side of the cut must contain roughly $n/2$ qubits, the partition problem becomes NP-hard; an $O(\sqrt{\log n})$-approximation is available via the SDP + metric-embedding technique of Arora-Rao-Vazirani~\cite{Arora2009Expander}. Because $\ell_i$ is a small constant (at most $4$ for most common two-qubit gates), the total number of tensor terms grows with the cut size~$m'$, i.e., $\prod_{i=1}^{m'}\ell_i = \ell$, not with the full Hilbert-space dimension $2^{n}$.

\begin{res}[Thread-level parallelism (TLP) criterion]
	Consider a workload whose objective value is
	\[
		\Tr\![\,O\,\Phi(\rho)],
		\qquad
		\Phi(\,\cdot\,)=
		\sum_{p=1}^{q}\
		\left(\sum_{i=1}^{m} c_{p,i}\,U_{p,i}\right)
		(\,\cdot\,)
		\left(\sum_{j=1}^{m} c_{p,j}\,U_{p,j}\right)^{\!\dagger},
	\]
	and assume
	\begin{enumerate}
		\item the data are tensor-separable,
		      $\rho=\bigotimes_{a=1}^{\mathcal A}\rho^{a}$ and
		      $O   =\bigotimes_{a=1}^{\mathcal A}O^{a}$;
		\item each $U_{p,i}$ is realized by a depth-$m_{d}$ circuit
		      $ U_{p,i}=%
			      \prod_{t=1}^{m_{d}}
			      \!(G^{(t)}_{\!e} \otimes_{a}U^{(t)}_{a})$
		      in which \emph{every} entangling layer factorizes as
		      \(      G^{(t)}_{\!e}=
		      \sum_{\alpha_t=1}^{\ell_t}
		      \bigotimes_{a=1}^{\mathcal A} g^{a}_{t\alpha_t}.
		      \)
	\end{enumerate}
	Then there exists a cut size $m' \le m_{d}$ such that the global
	expectation value factorizes into at most
	\[
		q \times\, m^{2} \times
		({\textstyle\prod_{t=1}^{m'}}\ell_t)^{2}
		\times\, \mathcal A
	\]
	independent single-subsystem traces. Each trace can be executed on an individual QPU node, and the final result is obtained by classical post-processing.
\end{res}

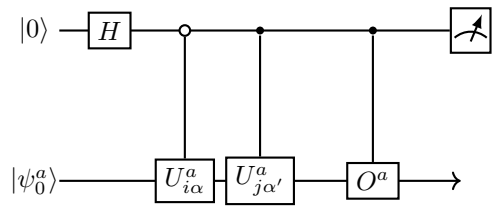
\begin{figure}[!h]
	\centering
	\begin{tikzpicture}[thick]
		\ctikzset{scale=2}
		\tikzstyle{every node}=[font=\normalsize,scale=1]
		\tikzstyle{operator} = [draw,shape=rectangle
		,fill=white,minimum width=1em, minimum height=1em]
		\tikzstyle{operator2} = [draw,shape=rectangle,fill=pink,minimum width=2.5em, minimum height=2.5em]
		\tikzstyle{operator22} = [draw,shape=rectangle,fill=white,minimum width=3em, minimum height=9.5em]
		\tikzstyle{operator3} = [draw,shape=rectangle,fill=white,minimum width=3em, minimum height=1em]
		\tikzstyle{operator4} = [draw,shape=rectangle,dashed, minimum width=1.5cm, minimum height=1cm]
		\tikzstyle{operator5} = [draw,shape=rectangle,dashed, minimum width=7.5cm, minimum height=4cm]
		\tikzstyle{operator6} = [draw=pink,shape=rectangle,dashed, minimum width=5cm, minimum height=4cm]
		\tikzstyle{phase} = [fill,shape=circle,minimum size=3pt,inner sep=0pt]
		\tikzstyle{surround} = [fill=blue!10,thick,draw=black,rounded corners=2mm]
		\tikzstyle{ellipsis} = [fill,shape=circle,minimum size=2pt,inner sep=0pt]
		\tikzstyle{ellipsis2} = [fill=white,draw=black, shape=circle,minimum size=4pt,inner sep=0.5pt]
		\tikzset{meter/.append style={fill=white, draw, inner sep=5, rectangle, font=\vphantom{A}, minimum width=15,
					path picture={\draw[black] ([shift={(.05,.2)}]path picture bounding box.south west) to[bend left=40] ([shift={(-.05,.2)}]path picture bounding box.south east);\draw[black,-latex] ([shift={(0,.15)}]path picture bounding box.south) -- ([shift={(.15,-.08)}]path picture bounding box.north);}}}
		\node at (0.,0) (qin){$\ket{0}$};
		\node at (0.,-1) (qin){$\ket{\psi _0^{a}}$};
		\node at (0.1,0) (q2){};
		\node[meter] (end3) at (2.9,0) {} edge [-] (q2);
		\node at (0.1,-1) (q3) {};
		\node[] (end3) at (2.9,-1) {} edge [<-] (q3);
		\node[operator] (op22) at (0.5,-0) {$H$} ;
		\node[ellipsis2] (cc1) at (1,-0) {};
		\node[operator] (op22) at (1,-1) {$U_{i\alpha}^{a}$} edge [-] (cc1);
		\node[phase] (cc1) at (1.5,-0) {};
		\node[operator] (op22) at (1.5,-1) {$U_{j\alpha'}^{a}$} edge [-] (cc1);
		\node[phase] (cc1) at (2.25,-0) {};
		\node[operator] (op22) at (2.25,-1) {$O^a$} edge [-] (cc1);
	\end{tikzpicture}
	\caption{Single-ancilla estimator: an ancillary qubit is used to estimate overlaps.}
	\label{fig:single_ancilla}
\end{figure}

\noindent\textbf{Implications.}
Whenever the input state and observable are already
tensor-separable, and all non-local gates admit a
known low-rank tensor expansion (e.g.\ \textsc{cnot}, \textsc{cz}),
the entire computation can be mapped onto a QPU cluster with one thread per tensor term-no quantum interconnect is required, and the required single-ancilla estimator is depicted as follows.
\begin{lem}[Single-ancilla estimator]
	\label{prop:single-ancilla}
	Let $\rho^{a}=\ket{\psi_0^{a}}\!\bra{\psi_0^{a}}$ be a pure state on
	sub-system~$a$, let
	$U^{a}_{i\alpha}$ and $U^{a}_{j\alpha'}$ be arbitrary unitaries on
	$\mathcal H_a$, and let $O^{a}$ be any unitary observable.
	The complex overlap
	\(  \Tr\!\bigl(    U^{a}_{i\alpha} \rho^{a}
	U^{a\,\dagger}_{j\alpha'}\,O^{a}
	\bigr)
	\)
	is obtained with a single ancillary qubit by
	\[
		\langle\sigma_{x}\rangle_{\rm anc}
		+
		i\,\langle\sigma_{y}\rangle_{\rm anc},
	\]
	where the expectation values are measured in the interferometric
	circuit, which is shown in Fig.~\ref{fig:single_ancilla}. Setting
	$j=\!i, \alpha'=\!\alpha$ recovers channel observables of the form
	$U_{i\alpha}\rho\,U_{i\alpha}^{\dagger}$.
\end{lem}

\textit{Proof sketch---}
Prepare the ancilla in $\ket{0}$ and subsystem $a$ in $\ket{\psi_{0}^{a}}$.
Apply the controlled unitary
\(
\ket{0}\!\bra{0}\!\otimes U^{a}_{i\alpha}
+
\ket{1}\!\bra{1}\!\otimes U^{a}_{j\alpha'}
\)
to obtain
\[
	\ket{\Psi^{(a)}_{i,j,\alpha,\alpha'}}=
	\tfrac{1}{\sqrt2}\bigl(
	\ket{0}\!\otimes\!U^{a}_{i\alpha}\ket{\psi_{0}^{a}}
	+\ket{1}\!\otimes\!U^{a}_{j\alpha'}\ket{\psi_{0}^{a}}
	\bigr).
\]
Next, insert $O^{a}$ only on the $\ket{1}$ branch:
\[
	\ket{\Psi^{O,(a)}_{i,j,\alpha, \alpha'}}=
	\tfrac{1}{\sqrt2}\bigl(
	\ket{0}\!\otimes\!U^{a}_{i\alpha}\ket{\psi_{0}^{a}}
	+\ket{1}\!\otimes\
	O^{a}U^{a}_{j\alpha'}\ket{\psi_{0}^{a}}
	\bigr).
\]
Tracing out subsystem $a$ leaves the ancilla density matrix
\[
	\rho_{\text{anc}}=
	\frac12
	\begin{pmatrix}
		1 &
		\bra{\psi_{0}^{a}}U^{a\,\dagger}_{i\alpha}O^{a}U^{a}_{j\alpha'}\ket{\psi_{0}^{a}} \\
		\bra{\psi_{0}^{a}}U^{a\,\dagger}_{j\alpha'}O^{a\,\dagger}U^{a}_{i\alpha}\ket{\psi_{0}^{a}}
		  & 1
	\end{pmatrix}.
\]
The off-diagonal element is precisely the desired overlap.
Projective measurements of the ancilla in the $\sigma_x$ and $\sigma_y$ bases read out its real and imaginary parts. For channels of the form $U_{i, \alpha}\rho U_{i, \alpha}^{\dagger}$ one simply sets $j, \alpha' =i, \alpha$, recovering the same protocol. \hfill\qedsymbol

Together, Lemma~\ref{prop:layered-decomp}
and~\ref{prop:single-ancilla} meet both prerequisites of
Eq.~\eqref{eq:parall}: an efficient factorization whose cost
depends only on the interaction cut, and a practical estimator
that requires just one extra qubit per subsystem.

\section{Supplementary information for configuration of current Hardware}
\label{app:hardware}

In this work, we implement clusters comprising either 8 or 16 benchtop NMR quantum processors, with each QPU node hosting two or three nuclear spin qubits.

First, we introduce a cluster of $8\times 3$-qubit QPUs based on the Triangulum platform, as detailed in~\cite{Feng2022SpinQ}. As shown in Fig.~\ref{fig:Fig2}(b), each device consists of three qubits, realized by the $^{19}$F nuclei in C$_2$F$_3$I molecules.

The relevant relaxation times, $T_1$ (longitudinal) and $T_2$ (transverse), are measured for all three nuclei and determine the system's coherence properties. The free evolution of each three-qubit NMR system is governed by the internal Hamiltonian:
\begin{eqnarray}\label{NMR_Hamiltonian3}
	\mathcal{H}_{\rm int} = \pi \nu_1 \sigma_z^1 + \pi \nu_2 \sigma_z^2 + \pi \nu_3 \sigma_z^3 + \frac{\pi}{2} J_{12} \sigma_z^1 \sigma_z^2 + \frac{\pi}{2} J_{23} \sigma_z^2 \sigma_z^3 + \frac{\pi}{2} J_{13} \sigma_z^3 \sigma_z^1,
\end{eqnarray}
where $\nu_i$ ($i=1,2,3$) denote the Larmor frequencies, and $J_{ij}$ are the $J$-coupling constants between the spins.

Universal quantum control is achieved by applying transverse radio-frequency (RF) pulses, described by
\begin{eqnarray}
	\label{NMR_RFHamiltonian3}
	\mathcal{H}_{\rm rf} = -\frac{1}{2} \omega_1 \sum_{i=1}^3 [\cos(\omega_{\rm rf} t + \phi)\sigma_x^{i} + \sin(\omega_{\rm rf} t + \phi)\sigma_y^i].
\end{eqnarray}
By tuning the RF field parameters-amplitude $\omega_1$, phase $\phi$, frequency $\omega_{\rm rf}$, and pulse duration-together with the system's internal evolution, arbitrary three-qubit quantum gates can be realized.

In our experiments, we employ 8 Triangulum units, with their main specifications summarized in Table~\ref{tab:triangulum_units}. Device-to-device variations in magnetic field strength, chemical shifts, and other parameters are hard to directly compensated via software calibration and by operating in the rotating frame. Therefore, variant circuit or shaped pulses are employed for robust control or generating standard quantum gates.

\begin{table}[!htbp]
	\centering
	\caption{Specifications of the SPINQ Triangulum Units}
	\small
	\renewcommand{\arraystretch}{1.1}
	\begin{tabular}{c c c c c}
		\toprule
		Model & Serial No. & Magnetic Field (T) & Homogeneity FWHM (ppm) & $^{19}$F Freq. (MHz) \\
		\midrule
		\multirow{8}{*}{Triangulum}
		      & T2023089   & 0.886              & 0.8                    & 35.470               \\
		      & T2023043   & 0.905              & 0.8                    & 36.249               \\
		      & T20250617  & 0.895              & 0.7                    & 35.858               \\
		      & T20250618  & 0.895              & 0.7                    & 35.858               \\
		      & T20231213  & 0.944              & 0.85                   & 37.809               \\
		      & T20250208  & 0.888              & 0.6                    & 35.581               \\
		      & T20250504  & 0.896              & 0.7                    & 35.875               \\
		      & T20250420  & 0.866              & 0.6                    & 34.684               \\
		\bottomrule
	\end{tabular}
	\label{tab:triangulum_units}
\end{table}

Second, we introduce $16\times 2$ qubit QPUs, which uses Gemini, referenced in~\cite{hou2021spinq}. As shown in Fig.~\ref{fig:Fig2}(b), the device comprises two qubits, represented by the nuclei $^1$H and$^{31}$P in Dimethylphosphite ((CH$_3$O)$_2$PH) molecules.
$T_1$ and $T_2$ represent the longitudinal and transverse relaxation times, respectively.
The free evolution of this $2$-qubit system is primarily governed by the internal Hamiltonian,
\begin{eqnarray}\label{NMR_Hamiltonian2}
	\mathcal{H}_{int}= \pi \nu _1  \sigma_z^1+\pi \nu _2  \sigma_z^2  + \frac{\pi}{2} J_0 \sigma_z^1 \sigma_z^2,
\end{eqnarray}
where $\nu_1$  and $\nu_2$ can be adjusted to $0$ Hz in a rotating frame, and $J_0 = 697.4$ Hz denotes the resonance frequency of the $J$-coupling strength between the spins. Further details are available in~\cite{hou2021spinq}. To control the system's evolution, transverse radio-frequency (r.f.) pulses serve as the control field, expressed as,
\begin{eqnarray}
	\label{NMR_RFHamiltonian}
	\mathcal{H}_{rf}=-\frac{1}{2} \sum_{i=1}^2 \omega_1^i(\cos(\omega_{rf} t+\phi^i)\sigma_x^{i}+\sin(\omega_{rf} t+\phi^i)\sigma_y^i).
\end{eqnarray}
By adjusting the parameters in the r.f. field [Eq.~\eqref{NMR_RFHamiltonian}], such as intensity $\omega_1$, phase $\phi$, frequency $\omega_{\text{rf}}$, and duration, the theoretical achievement of two-qubit universal quantum gates is possible through the combination of the system's internal dynamics.

In our experiments, we employ 16 Gemini units, with their main specifications summarized in Table~\ref{tab:gemini_units}. While there are device-to-device variations in magnetic field strength, chemical shifts, and other parameters, these differences are compensated in software by appropriate calibration and by working in the rotating frame. Standard NMR techniques such as hard and shaped pulses are employed for robust control.

\begin{table}[!htbp]
	\centering
	\caption{Specifications of the SPINQ Gemini Lab Units}
	\small
	\renewcommand{\arraystretch}{1.1}
	\begin{tabular}{c c c c c c}
		\toprule
		Model & Serial No. & Magnetic Field (T) & Homogeneity FWHM (ppm) & $^1$H Freq. (MHz) & $^{31}$P Freq. (MHz) \\
		\midrule
		\multirow{16}{*}{Gemini Lab}
		      & L20250104  & 0.654              & 0.6                    & 27.848            & 11.273               \\
		      & L20250607  & 0.644              & 0.5                    & 27.436            & 11.106               \\
		      & L20250415  & 0.661              & 0.3                    & 28.138            & 11.390               \\
		      & L20250414  & 0.657              & 0.5                    & 27.984            & 11.328               \\
		      & L20250419  & 0.656              & 0.4                    & 27.949            & 11.314               \\
		      & L20250413  & 0.660              & 0.3                    & 28.097            & 11.375               \\
		      & L20250412  & 0.661              & 0.4                    & 28.124            & 11.385               \\
		      & L20250418  & 0.645              & 0.4                    & 27.474            & 11.121               \\
		      & L20250417  & 0.653              & 1.0                    & 27.806            & 11.256               \\
		      & L20250411  & 0.638              & 0.4                    & 27.172            & 10.999               \\
		      & L20250410  & 0.644              & 0.5                    & 27.405            & 11.094               \\
		      & L20250602  & 0.651              & 0.7                    & 27.721            & 11.221               \\
		      & L20250421  & 0.636              & 0.6                    & 27.061            & 10.954               \\
		      & L20250416  & 0.654              & 0.5                    & 27.865            & 11.280               \\
		      & L20250620  & 0.648              & 0.5                    & 27.578            & 11.163               \\
		      & L20250606  & 0.621              & 0.7                    & 26.449            & 10.706               \\
		\bottomrule
	\end{tabular}
	\label{tab:gemini_units}
\end{table}

\section{Supplementary information for GHZ state preparation}
\label{app:exp1}
\subsection{Our method}
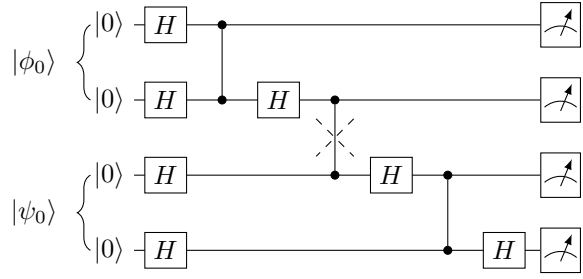
\begin{figure}[h]
	\centering
	\begin{tikzpicture}[scale=1, every node/.style={scale=1}]
		\tikzstyle{gate} = [draw, fill=white, minimum size=1.2em]
		\tikzstyle{control} = [draw, fill=black, shape=circle, minimum size=3pt, inner sep=0pt]
		\tikzstyle{control2} = [draw, fill=white, shape=circle, minimum size=3pt, inner sep=0pt]
		\tikzset{meter/.append style={fill=white, draw, inner sep=5, rectangle, font=\vphantom{A}, minimum width=15,
					path picture={\draw[black] ([shift={(.05,.2)}]path picture bounding box.south west) to[bend left=40] ([shift={(-.05,.2)}]path picture bounding box.south east);\draw[black,-latex] ([shift={(0,.15)}]path picture bounding box.south) -- ([shift={(.15,-.08)}]path picture bounding box.north);}}}
		\node at (0.5,0) (q1){$\ket{0}$} ;
		\node[meter] (end1) at (6.5,0 ) {} edge [-] (q1);
		\node at (0.5,-1) (q2){$\ket{0}$} ;
		\node[meter] (end2) at (6.5,-1) {} edge [-] (q2);
		\node at (0.5,-2) (q3){$\ket{0}$} ;
		\node[meter] (end2) at (6.5,-2) {} edge [-] (q3);
		\node at (0.5,-3) (q4){$\ket{0}$} ;
		\node[meter] (end2) at (6.5,-3) {} edge [-] (q4);
		\draw [decorate,decoration={brace,amplitude=5pt,mirror}] (0.25, 0) -- (0.25, -1);
		\node at (-0.5,-0.5) (q4){$\ket{\phi_0}$};
		\node at (-0.5,-2.5) (q4){$\ket{\psi_0}$};
		\draw [decorate,decoration={brace,amplitude=5pt,mirror}] (0.25, -2) -- (0.25, -3);
		\draw (1.25,0) node[gate] (H1) {$H$};
		\draw (1.25,-1) node[gate] (H1) {$H$};
		\draw (1.25,-2) node[gate] (H1) {$H$};
		\draw (1.25,-3) node[gate] (H1) {$H$};
		\draw (2,0) node[control] (C) {};
		\draw (C) -- (2.,-1);
		\draw (2.,-1) node[control] (X1) {};
		\draw (2.75,-1) node[gate] (H1) {$H$};
		\draw (3.5,-1) node[control] (C) {};
		\draw (C) -- (3.5,-2);
		\draw (3.5,-2) node[control] (X2) {};
		\draw (4.25,-2) node[gate] (H1) {$H$};
		\draw (5,-2) node[control] (C) {};
		\draw (C) -- (5,-3);
		\draw (5,-3) node[control] (X2) {};
		\draw (5.75,-3) node[gate] (H1) {$H$};
		\draw[dashed] (3.75,-1.25) -- (3.25,-1.75);
		\draw[dashed] (3.25,-1.25) -- (3.75,-1.75);
	\end{tikzpicture}
	\caption{Typical quantum circuit for 4 qubit GHZ state, where control-Z gate between the 2nd and 3rd qubits is cut to generate sub-circuits.}
	\label{fig:ghz}
\end{figure}

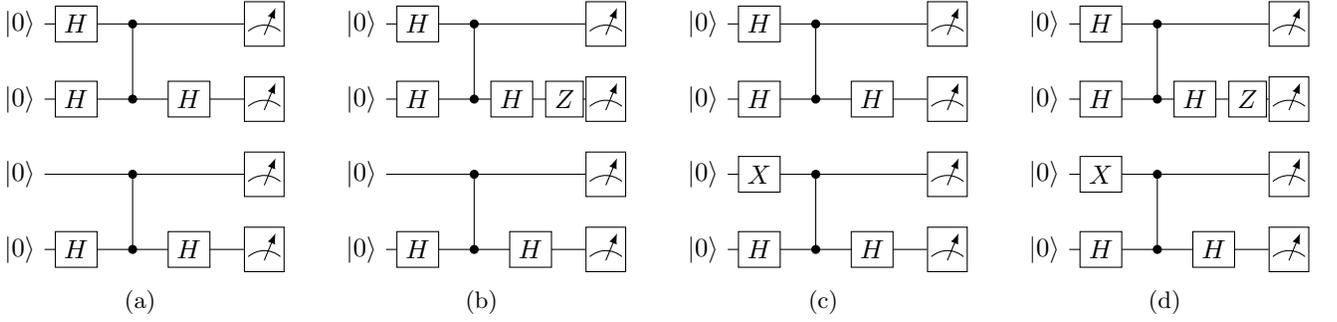
\begin{figure}[h]
	\centering
	\tikzstyle{gate}     = [draw, fill=white, minimum size=1.2em]
	\tikzstyle{control}  = [draw, fill=black, shape=circle, minimum size=3pt, inner sep=0pt]
	\tikzstyle{control2} = [draw, fill=white, shape=circle, minimum size=3pt, inner sep=0pt]
	\tikzset{meter/.append style={
				fill=white, draw, inner sep=5, rectangle, minimum width=15,
				path picture={
						\draw[black] ([shift={(.05,.2)}]path picture bounding box.south west)
						to[bend left=40] ([shift={(-.05,.2)}]path picture bounding box.south east);
						\draw[black,-latex] ([shift={(0,.15)}]path picture bounding box.south)
						-- ([shift={(.15,-.08)}]path picture bounding box.north);
					}
			}}

	\begin{subfigure}[b]{0.24\textwidth}
		\centering
		\begin{tikzpicture}[scale=1, every node/.style={scale=1}]
			\tikzstyle{gate} = [draw, fill=white, minimum size=1.2em]
			\tikzstyle{control} = [draw, fill=black, shape=circle, minimum size=3pt, inner sep=0pt]
			\tikzstyle{control2} = [draw, fill=white, shape=circle, minimum size=3pt, inner sep=0pt]
			\tikzset{meter/.append style={fill=white, draw, inner sep=5, rectangle, font=\vphantom{A}, minimum width=15,
						path picture={\draw[black] ([shift={(.05,.2)}]path picture bounding box.south west) to[bend left=40] ([shift={(-.05,.2)}]path picture bounding box.south east);\draw[black,-latex] ([shift={(0,.15)}]path picture bounding box.south) -- ([shift={(.15,-.08)}]path picture bounding box.north);}}}
			\node at (0.5,0) (q1){$\ket{0}$} ;
			\node[meter] (end1) at (3.75,0 ) {} edge [-] (q1);
			\node at (0.5,-1) (q2){$\ket{0}$} ;
			\node[meter] (end2) at (3.75,-1) {} edge [-] (q2);
			\node at (0.5,-2) (q3){$\ket{0}$} ;
			\node[meter] (end2) at (3.75,-2) {} edge [-] (q3);
			\node at (0.5,-3) (q4){$\ket{0}$} ;
			\node[meter] (end2) at (3.75,-3) {} edge [-] (q4);
			\draw (1.25,0) node[gate] (H1) {$H$};
			\draw (1.25,-1) node[gate] (H1) {$H$};
			\draw (1.25,-3) node[gate] (H1) {$H$};
			\draw (2,0) node[control] (C) {};
			\draw (C) -- (2.,-1);
			\draw (2.,-1) node[control] (X1) {};
			\draw (2.75,-1) node[gate] (H1) {$H$};

			\draw (2,-2) node[control] (C) {};
			\draw (C) -- (2,-3);
			\draw (2,-3) node[control] (X2) {};
			\draw (2.75,-3) node[gate] (H1) {$H$};
		\end{tikzpicture}
		\caption{}
	\end{subfigure}%
	\hfill
	\begin{subfigure}[b]{0.24\textwidth}
		\centering
		\begin{tikzpicture}[scale=1, every node/.style={scale=1}]
			\tikzstyle{gate} = [draw, fill=white, minimum size=1.2em]
			\tikzstyle{control} = [draw, fill=black, shape=circle, minimum size=3pt, inner sep=0pt]
			\tikzstyle{control2} = [draw, fill=white, shape=circle, minimum size=3pt, inner sep=0pt]
			\tikzset{meter/.append style={fill=white, draw, inner sep=5, rectangle, font=\vphantom{A}, minimum width=15,
						path picture={\draw[black] ([shift={(.05,.2)}]path picture bounding box.south west) to[bend left=40] ([shift={(-.05,.2)}]path picture bounding box.south east);\draw[black,-latex] ([shift={(0,.15)}]path picture bounding box.south) -- ([shift={(.15,-.08)}]path picture bounding box.north);}}}
			\node at (0.5,0) (q1){$\ket{0}$} ;
			\node[meter] (end1) at (3.75,0 ) {} edge [-] (q1);
			\node at (0.5,-1) (q2){$\ket{0}$} ;
			\node[meter] (end2) at (3.75,-1) {} edge [-] (q2);
			\node at (0.5,-2) (q3){$\ket{0}$} ;
			\node[meter] (end2) at (3.75,-2) {} edge [-] (q3);
			\node at (0.5,-3) (q4){$\ket{0}$} ;
			\node[meter] (end2) at (3.75,-3) {} edge [-] (q4);
			\draw (1.25,0) node[gate] (H1) {$H$};
			\draw (1.25,-1) node[gate] (H1) {$H$};
			\draw (1.25,-3) node[gate] (H1) {$H$};
			\draw (2,0) node[control] (C) {};
			\draw (C) -- (2.,-1);
			\draw (2.,-1) node[control] (X1) {};
			\draw (2.5,-1) node[gate] (H1) {$H$};
			\draw (3.2,-1) node[gate] (H1) {$Z$};
			\draw (2,-2) node[control] (C) {};
			\draw (C) -- (2,-3);
			\draw (2,-3) node[control] (X2) {};
			\draw (2.75,-3) node[gate] (H1) {$H$};
		\end{tikzpicture}
		\caption{}
	\end{subfigure}%
	\hfill
	\begin{subfigure}[b]{0.24\textwidth}
		\centering
		\begin{tikzpicture}[scale=1, every node/.style={scale=1}]
			\tikzstyle{gate} = [draw, fill=white, minimum size=1.2em]
			\tikzstyle{control} = [draw, fill=black, shape=circle, minimum size=3pt, inner sep=0pt]
			\tikzstyle{control2} = [draw, fill=white, shape=circle, minimum size=3pt, inner sep=0pt]
			\tikzset{meter/.append style={fill=white, draw, inner sep=5, rectangle, font=\vphantom{A}, minimum width=15,
						path picture={\draw[black] ([shift={(.05,.2)}]path picture bounding box.south west) to[bend left=40] ([shift={(-.05,.2)}]path picture bounding box.south east);\draw[black,-latex] ([shift={(0,.15)}]path picture bounding box.south) -- ([shift={(.15,-.08)}]path picture bounding box.north);}}}
			\node at (0.5,0) (q1){$\ket{0}$} ;
			\node[meter] (end1) at (3.75,0 ) {} edge [-] (q1);
			\node at (0.5,-1) (q2){$\ket{0}$} ;
			\node[meter] (end2) at (3.75,-1) {} edge [-] (q2);
			\node at (0.5,-2) (q3){$\ket{0}$} ;
			\node[meter] (end2) at (3.75,-2) {} edge [-] (q3);
			\node at (0.5,-3) (q4){$\ket{0}$} ;
			\node[meter] (end2) at (3.75,-3) {} edge [-] (q4);
			\draw (1.25,0) node[gate] (H1) {$H$};
			\draw (1.25,-1) node[gate] (H1) {$H$};
			\draw (1.25,-3) node[gate] (H1) {$H$};
			\draw (2,0) node[control] (C) {};
			\draw (C) -- (2.,-1);
			\draw (2.,-1) node[control] (X1) {};
			\draw (2.75,-1) node[gate] (H1) {$H$};

			\draw (2,-2) node[control] (C) {};
			\draw (C) -- (2,-3);
			\draw (2,-3) node[control] (X2) {};
			\draw (1.25,-2) node[gate] (H1) {$X$};
			\draw (2.75,-3) node[gate] (H1) {$H$};
		\end{tikzpicture}
		\caption{}
	\end{subfigure}%
	\hfill
	\begin{subfigure}[b]{0.24\textwidth}
		\centering
		\begin{tikzpicture}[scale=1, every node/.style={scale=1}]
			\tikzstyle{gate} = [draw, fill=white, minimum size=1.2em]
			\tikzstyle{control} = [draw, fill=black, shape=circle, minimum size=3pt, inner sep=0pt]
			\tikzstyle{control2} = [draw, fill=white, shape=circle, minimum size=3pt, inner sep=0pt]
			\tikzset{meter/.append style={fill=white, draw, inner sep=5, rectangle, font=\vphantom{A}, minimum width=15,
						path picture={\draw[black] ([shift={(.05,.2)}]path picture bounding box.south west) to[bend left=40] ([shift={(-.05,.2)}]path picture bounding box.south east);\draw[black,-latex] ([shift={(0,.15)}]path picture bounding box.south) -- ([shift={(.15,-.08)}]path picture bounding box.north);}}}
			\node at (0.5,0) (q1){$\ket{0}$} ;
			\node[meter] (end1) at (3.75,0 ) {} edge [-] (q1);
			\node at (0.5,-1) (q2){$\ket{0}$} ;
			\node[meter] (end2) at (3.75,-1) {} edge [-] (q2);
			\node at (0.5,-2) (q3){$\ket{0}$} ;
			\node[meter] (end2) at (3.75,-2) {} edge [-] (q3);
			\node at (0.5,-3) (q4){$\ket{0}$} ;
			\node[meter] (end2) at (3.75,-3) {} edge [-] (q4);
			\draw (1.25,0) node[gate] (H1) {$H$};
			\draw (1.25,-1) node[gate] (H1) {$H$};
			\draw (1.25,-3) node[gate] (H1) {$H$};
			\draw (2,0) node[control] (C) {};
			\draw (C) -- (2.,-1);
			\draw (2.,-1) node[control] (X1) {};
			\draw (2.5,-1) node[gate] (H1) {$H$};
			\draw (3.2,-1) node[gate] (H1) {$Z$};
			\draw (1.25,-2) node[gate] (H1) {$X$};
			\draw (2,-2) node[control] (C) {};
			\draw (C) -- (2,-3);
			\draw (2,-3) node[control] (X2) {};
			\draw (2.75,-3) node[gate] (H1) {$H$};
		\end{tikzpicture}
		\caption{}
	\end{subfigure}

	\caption{Experimental circuit for 4-qubit GHZ state. circuits in (a-d) are generated by cutting the control-Z gate between the 2nd and 3rd qubits and replace it by a sum of local operations. Those circuit are summed in a superposition way.}
	\label{fig:eq_ghz_4x1}
\end{figure}

Fig.~\ref{fig:ghz} provides circuit to generate a 4-qubit GHZ state using a 4 qubit unitary transformation.
According to Lemma~\ref{prop:layered-decomp}, the 4-qubit GHZ circuit is decomposed into parallel two-qubit subcircuits, each suitable for small-scale hardware.
As the CNOT gate between qubits 2 and 3 can be ``cut'' and replaced by a sum of local operations:
\[
	U_{cnot} = \frac{1}{2} \left(\openone \otimes \openone + \sigma_z \otimes \openone + \openone \otimes \sigma_x - \sigma_z \otimes \sigma_x\right),
\]
with Pauli operators $\sigma_{x,y,z}$ and identity $\openone$, we can generate a replaced circuits, shown in Fig.~\ref{fig:eq_ghz_4x1}.

\begin{figure}[h]
	\centering
	\begin{subfigure}[b]{0.48\textwidth}
		\centering
		\begin{tikzpicture}[thick]
			\ctikzset{scale=2}
			\tikzstyle{every node}=[font=\normalsize,scale=1]
			\tikzstyle{operator} = [draw,shape=rectangle
			,fill=white,minimum width=1em, minimum height=1em]
			\tikzstyle{operator2} = [draw,shape=rectangle,fill=pink,minimum width=2.5em, minimum height=5em]
			\tikzstyle{operator22} = [draw,shape=rectangle,fill=white,minimum width=3em, minimum height=9.5em]
			\tikzstyle{operator3} = [draw,shape=rectangle,fill=white,minimum width=3em, minimum height=1em]
			\tikzstyle{operator4} = [draw,shape=rectangle,dashed, minimum width=1.5cm, minimum height=1cm]
			\tikzstyle{operator5} = [draw,shape=rectangle,dashed, minimum width=7.5cm, minimum height=4cm]
			\tikzstyle{operator6} = [draw=pink,shape=rectangle,dashed, minimum width=5cm, minimum height=4cm]
			\tikzstyle{phase} = [fill,shape=circle,minimum size=3pt,inner sep=0pt]
			\tikzstyle{surround} = [fill=blue!10,thick,draw=black,rounded corners=2mm]
			\tikzstyle{ellipsis} = [fill,shape=circle,minimum size=2pt,inner sep=0pt]
			\tikzstyle{ellipsis2} = [fill=white,draw=black, shape=circle,minimum size=6pt,inner sep=0.5pt]
			\tikzset{meter/.append style={fill=white, draw, inner sep=5, rectangle, font=\vphantom{A}, minimum width=15,
						path picture={\draw[black] ([shift={(.05,.2)}]path picture bounding box.south west) to[bend left=40] ([shift={(-.05,.2)}]path picture bounding box.south east);\draw[black,-latex] ([shift={(0,.15)}]path picture bounding box.south) -- ([shift={(.15,-.08)}]path picture bounding box.north);}}}
			\node at (0.,0) (qin){$\ket{0}$};
			\node at (0.,-1) (qin){$\ket{0}$};
			\node at (0.,-1.5) (qin){$\ket{0}$};
			\node at (0.1,0) (q2){};
			\node[meter] (end3) at (2.9,0) {} edge [-] (q2);
			\node at (0.1,-1) (q3) {};
			\node[] (end3) at (2.9,-1) {} edge [<-] (q3);
			\node at (0.1,-1.5) (q3) {};
			\node[] (end3) at (2.9,-1.5) {} edge [<-] (q3);
			\node[operator] (op22) at (0.5,-0) {$H$} ;
			\node[operator] (op22) at (0.5,-1) {$H$} ;
			\node[operator] (op22) at (0.5,-1.5) {$H$} ;
			\node[phase] (cc1) at (0.75,-1) {};
			\node[phase] (cc2) at (0.75,-1.5) {}edge [-] (cc1);
			\node[operator] (op22) at (1,-1.5) {$H$} ;
			\node[phase] (cc1) at (1.5,-0) {};
			\node[operator] (op22) at (1.5,-1.5) {$Z$} edge [-] (cc1);
			\node[phase] (cc1) at (2.25,-0) {};
			\node[operator2] (op22) at (2.25,-1.25) {} edge [-] (cc1);
			\node at (2.25,-1.25) (qtex){\textcolor{black}{$M$}};
		\end{tikzpicture}
		\caption{}
	\end{subfigure}%
	\hfill
	\begin{subfigure}[b]{0.48\textwidth}
		\centering
		\begin{tikzpicture}[thick]
			\ctikzset{scale=2}
			\tikzstyle{every node}=[font=\normalsize,scale=1]
			\tikzstyle{operator} = [draw,shape=rectangle
			,fill=white,minimum width=1em, minimum height=1em]
			\tikzstyle{operator2} = [draw,shape=rectangle,fill=pink,minimum width=2.5em, minimum height=5em]
			\tikzstyle{operator22} = [draw,shape=rectangle,fill=white,minimum width=3em, minimum height=9.5em]
			\tikzstyle{operator3} = [draw,shape=rectangle,fill=white,minimum width=3em, minimum height=1em]
			\tikzstyle{operator4} = [draw,shape=rectangle,dashed, minimum width=1.5cm, minimum height=1cm]
			\tikzstyle{operator5} = [draw,shape=rectangle,dashed, minimum width=7.5cm, minimum height=4cm]
			\tikzstyle{operator6} = [draw=pink,shape=rectangle,dashed, minimum width=5cm, minimum height=4cm]
			\tikzstyle{phase} = [fill,shape=circle,minimum size=3pt,inner sep=0pt]
			\tikzstyle{surround} = [fill=blue!10,thick,draw=black,rounded corners=2mm]
			\tikzstyle{ellipsis} = [fill,shape=circle,minimum size=2pt,inner sep=0pt]
			\tikzstyle{ellipsis2} = [fill=white,draw=black, shape=circle,minimum size=6pt,inner sep=0.5pt]
			\tikzset{meter/.append style={fill=white, draw, inner sep=5, rectangle, font=\vphantom{A}, minimum width=15,
						path picture={\draw[black] ([shift={(.05,.2)}]path picture bounding box.south west) to[bend left=40] ([shift={(-.05,.2)}]path picture bounding box.south east);\draw[black,-latex] ([shift={(0,.15)}]path picture bounding box.south) -- ([shift={(.15,-.08)}]path picture bounding box.north);}}}
			\node at (0,0) (qin){$\ket{0}$};
			\node at (0,-1) (qin){$\ket{0}$};
			\node at (0,-1.5) (qin){$\ket{0}$};
			\node at (0.1,0) (q2){};
			\node[meter] (end3) at (2.9,0) {} edge [-] (q2);
			\node at (0.1,-1) (q3) {};
			\node[] (end3) at (2.9,-1) {} edge [<-] (q3);
			\node at (0.1,-1.5) (q3) {};
			\node[] (end3) at (2.9,-1.5) {} edge [<-] (q3);
			\node[operator] (op22) at (0.5,-0) {$H$} ;
			\node[operator] (op22) at (1.25,-1) {$H$} ;
			\node[operator] (op22) at (1.25,-1.5) {$H$} ;
			\node[phase] (cc1) at (1.5,-1) {};
			\node[phase] (cc2) at (1.5,-1.5) {}edge [-] (cc1);
			\node[operator] (op22) at (1.75,-1.5) {$H$} ;
			\node[phase] (cc1) at (0.75,-0) {};
			\node[operator] (op22) at (0.75,-1) {$X$} edge [-] (cc1);
			\node[phase] (cc1) at (2.25,-0) {};
			\node[operator2] (op22) at (2.25,-1.25) {} edge [-] (cc1);
			\node at (2.25,-1.25) (qtex){\textcolor{black}{$M$}};
		\end{tikzpicture}
		\caption{}
	\end{subfigure}%
	\caption{Experimental circuit for 4-qubit GHZ state. (a) and (b) simulate the overlaps induced via $\ket{\phi_j}$ and $\ket{\psi_k}$, respectably. By varying $M$, we can get complete information on 4-qubit GHZ state.}
	\label{fig:exp_ghz}
\end{figure}
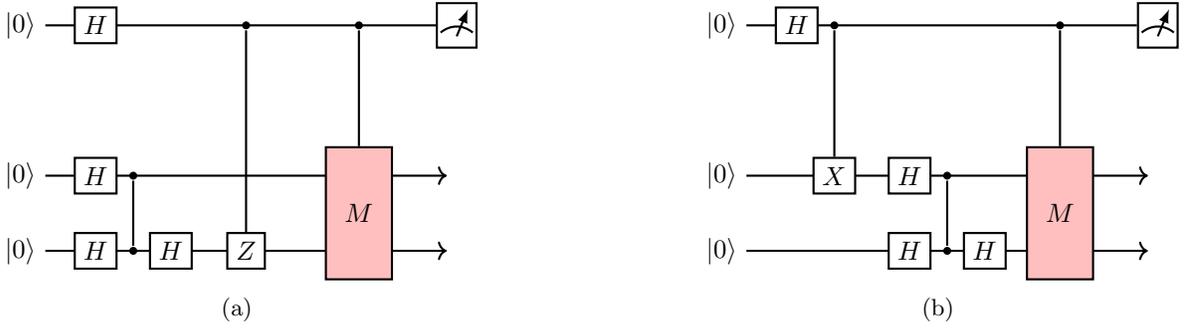

Mathematically, the output state is a sum of product of sub-states
\[\sum_{\{j,k\}}(-1)^{j-1 \cdot k-1}\ket{\phi_j}\otimes \ket{\psi_k}, \]
where $j,k = 1, 2$. $\ket{\phi_1}=U_{cnot} (H\otimes I) \ket{00}$, $\ket{\phi_2}=I\otimes \sigma_z \ket{\phi_1}$, $\ket{\psi_1}=U_{cnot} \ket{00}$, and  $\ket{\psi_2}=U_{cnot}(\sigma_x\otimes I) \ket{00}$.
This indicates the entire preparation process can be parallelized.
Via combining single-ancilla estimator, we can form the experimental circuit depicted in Fig.~\ref{fig:exp_ghz}, and measure the ancilla qubit with $\sigma_x$ and $\sigma_y$. Through varying $M$ in the circuit for all two-qubit Pauli bases, we can get sufficient information for reconstructing a GHZ state.

In this experiment, two types of circuits are employed, with $M$ varied to cover all Pauli bases required for full tomography. If every expectation value for a given Pauli basis is counted as an independent measurement, a total of $4\times16$ measurements are required. This is notably reduced compared to conventional four-qubit experiments, as both real and imaginary parts are separately recorded using the ancilla-assisted circuit.

\begin{figure}[!ht]
	\centering
	\includegraphics[width=1\linewidth]{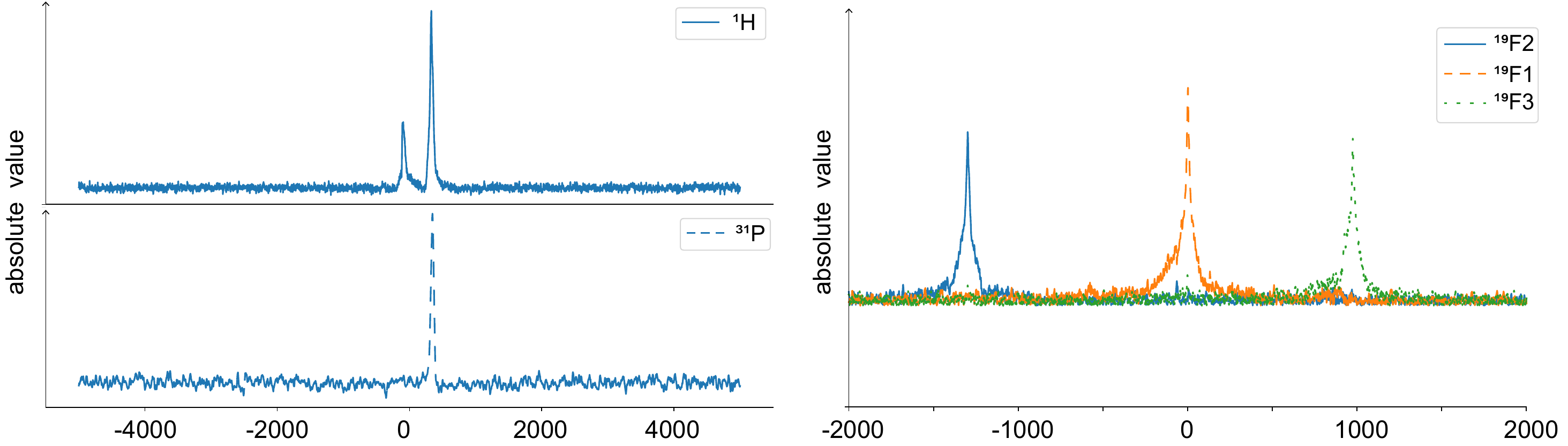}
	\caption{Spectrum of all spins when system is at psudo-pure state for (CH$_3$O)$_2$PHO (left) and C$_2$F$_3$I (right). The horizontal axis is the chemical shift while vertical axis is absolute value of signal.}
	\label{fig:pps_spec}
\end{figure}
Each node of the Triangulum-pro desktop spectrometer hosts three $^{19}\mathrm F$ nuclear spins in a C$_2$F$_3$I molecule, with typical coherence times of $T_2\approx500~ms$ and a calibrated maximum Rabi rate of $\lvert\omega_1\rvert/2\pi = 8.3 ~{kHz}$. Eight such nodes are synchronized by a common clock and exchange shot data via Gigabit Ethernet. We focus here on one representative experiment to illustrate the procedure:

\textit{Initialization ---} A spatial operation sequence is applied to transform the thermal equilibrium state into a pseudo-pure ground state $\ket{000}$. The resulting state fidelity reaches $98.5\%$ (see spectrum in Fig.~\ref{fig:pps_spec}).

\textit{Gate Implementation:}
The unitary operator represented by the sub-circuits (shown in Figure~\ref{fig:exp_ghz} dashed boxes) is realized by a single shaped pulse of $28~ms$ duration, discretized into $M=800$ time slices ($\Delta t = 35~\mu$s).

\textit{Pulse Optimization---}
An initial guess is generated using random spline envelopes $\{B_x(t), B_y(t)\}$ constrained by $\lvert B\rvert \le 0.4\lvert\omega_1\rvert$, and accepted only if the simulated unitary fidelity exceeds $20\%$. The GRAPE algorithm is then employed to maximize the worst-case channel fidelity $\mathcal{F}_{\mathrm{min}}$ across nine error configurations, spanning three static $B_0$ offsets ($-20$, $0$, and $+20$ Hz) and three RF amplitude miscalibrations (0.95, 1.00, 1.05). Optimization typically converges within $15$--$20$ L-BFGS iterations. The final control amplitudes and phases are discretized to $0.01\%$ resolution and exported as a waveform file with total pulse duration of $28~\mathrm{ms}$. GRAPE simulations yield an average gate fidelity of $\overline{\mathcal{F}}_{\mathrm{gate}} = 99.3\%$ over all error conditions.

\textit{GHZ State Preparation --- } Finally, the GRAPE-compiled pulse is executed concurrently on all nodes in the cluster. Preparing GHZ State across two neighboring nodes yields a state fidelity of $ 93.8\%$.

\subsection{Quantum Circuit Cutting}
Similarly, we implement quantum circuit cutting to decompose a 4-qubit GHZ preparation circuit into a sum of eleven 2-qubit subcircuits, following the procedure of~\cite{mitarai2021constructing}. Unlike our architecture, which allows coherent superpositions and therefore requires ancillary systems for coherent control, quantum circuit cutting operates entirely classically. That is, the summation in circuit cutting is incoherent-no entanglement or interference is preserved-thereby avoiding the need for ancilla but increasing the classical processing overhead.

Quantum circuit cutting is a powerful strategy to simulate large quantum circuits on smaller hardware by dividing a global circuit into independently executable fragments. This technique is particularly valuable for near-term quantum devices constrained by limited qubit counts, coherence times, and gate fidelities. It enables logical circuit execution beyond physical limitations, relying on classical coordination and tomography-based post-processing.

\begin{figure}[!ht]
	\centering
	\begin{tikzpicture}[baseline=(current bounding box.center), scale=1.0]
		\tikzstyle{gate1} = [draw, fill=white, rectangle, minimum width=1.2em, minimum height= 3.5em]
		\tikzstyle{gate2} = [draw, fill=white, minimum size=1.2em]
		\draw[thick] (0,0) -- (1.2,0);
		\draw[thick] (0,-1) -- (1.2,-1);
		\node at (0.1,-0) (){$/$};
		\node at (0.1,-1) (){$/$};
		\draw (0.6,-0.5) node[gate1] (X3) {$U_{ab}$};
		\node at (2.2,-0.5) (){$= \sum_i c_i $};
		\draw[thick] (3,0) -- (5,0);
		\draw[thick] (3,-1) -- (5,-1);
		\node at (3.2,-0) (){$/$};
		\node at (3.2,-1) (){$/$};
		\draw (4,0) node[gate2] (X1) {$U_{ai}$};
		\draw (4,-1) node[gate2] (X2) {$U_{bi}$};
	\end{tikzpicture}
	\caption{Schematic of quantum circuit cutting from an entire system to a bi-partition one.}
	\label{sfig:qcc}
\end{figure}
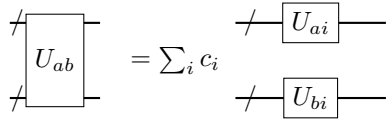

In the spatial (or state-space) variant of circuit cutting, we approximate a two-subsystem unitary channel as
\begin{equation}\label{U_appro2}
	U_{ab} \rho U_{ab}^{\dagger}
	\approx
	\sum_{i} c_i
	U_{ai}\rho_a U_{ai}^{\dagger} \otimes
	U_{bi}\rho_b U_{bi}^{\dagger},
\end{equation}
assuming an initially factorized input state $\rho=\rho_a\otimes\rho_b$.
This expansion allows the joint evolution under $U_{ab}$ to be mimicked by local operations $U_{ai}$ and $U_{bi}$ on separate devices, followed by classical post-processing. Figure~\ref{sfig:qcc} illustrates this decomposition, where the bipartite gate is expressed as a convex combination of local unitaries applied independently to subsystems $a$ and $b$.

In our demonstration, we apply this scheme to the 4-qubit GHZ circuit shown in Fig.~\ref{fig:ghz}. The key component-namely, the controlled-Z gate connecting the two partitions-is decomposed into a sum over tensor products of single-qubit operators. This decomposition reduces the circuit into 2-qubit subcircuits, which can then be distributed across multiple 2-qubit quantum processors.

To reconstruct the final GHZ state, each subcircuit is executed separately and its output measured in all required Pauli bases. In total, 10 distinct subcircuits are used, and each requires measurements in 16 different Pauli settings for full state tomography. Therefore, a total of $10 \times 16 = 160$ measurement configurations are needed for complete reconstruction.

\begin{figure}[!ht]
	\centering
	\begin{tikzpicture}[scale=1, thick]
		\ctikzset{scale=1}
		\tikzstyle{every node}=[font=\normalsize,scale=1]
		\tikzstyle{gate2} = [draw, fill=white, minimum size=1.em]
		\node at (0,-0.5) {$\frac{1}{2}$};
		\draw[thick] (0.25,0) -- (2.25,0);
		\draw[thick] (0.25,-1) -- (2.25,-1);
		\draw (1.25,0) node [gate2] (X1) {$e^{i\pi Z/4}$};
		\draw (1.25,-1) node [gate2] (X2) {$e^{i\pi Z/4}$};
		\tikzstyle{gate2} = [draw, fill=white, minimum size=1.em]
		\node at (2.75,-0.5) {$+\frac{1}{2}$};
		\draw[thick] (3,0) -- (5,0);
		\draw[thick] (3,-1) -- (5,-1);
		\draw (4,0) node [gate2] (X1) {$e^{-i\pi Z/4}$};
		\draw (4,-1) node [gate2] (X2) {$e^{-i\pi Z/4}$};
		\node at (6.5,-0.6) {$-\frac{1}{2}\sum_{\alpha \in \{\pm 1\}^{2}} \alpha_1 \alpha_2$};
		\draw [decorate,decoration={brace,amplitude=5pt,mirror}] (8, 0.2) -- (8, -1.2);
		\draw[thick] (8,0) -- (10.5,0);
		\draw[thick] (8,-1) -- (10.5,-1);
		\draw (9.25,0) node [gate2] (X1) {$(I+\alpha_1Z)/2$};
		\draw (9.25,-1) node [gate2] (X2) {$e^{i(\alpha_2+1)\pi Z/4}$};
		\node at (10.75,-0.5) {$-$};
		\draw[thick] (11,0) -- (13.5,0);
		\draw[thick] (11,-1) -- (13.5,-1);
		\draw (12.25,0) node [gate2] (X1) {$e^{i(\alpha_1+1)\pi Z/4}$};
		\draw (12.25,-1) node [gate2] (X2) {$(I+\alpha_2Z)/2$};
		\draw [decorate,decoration={brace,amplitude=3pt}] (13.6, 0.2) -- (13.6, -1.2);
	\end{tikzpicture}
	\caption{Decompositions of controlled-Z gate into a sequence of single-qubit operations.}
	\label{fig:decomposition}
\end{figure}

Each node of the Gemini-Lab desktop spectrometer hosts a $^{1}\mathrm H$ nuclear spin and a  $^{31}\mathrm P$ nuclear spins in a Dimethyl Phosphite molecule (CH$_4$O)$_2$PH, with typical coherence times of $T_2\approx500~ms$ and a calibrated maximum Rabi rate of $\lvert\omega_1\rvert/2\pi = 6.25 ~{kHz}$. Eight such nodes are synchronized by a common clock and exchange shot data via Gigabit Ethernet. We focus here on one representative experiment to illustrate the procedure:

\textit{Initialization ---} An operation sequence named relaxation method is applied to transform the thermal equilibrium state into a pseudo-pure ground state $\ket{00}$. The resulting state fidelity reaches $99.5\%$ (See Fig.~\ref{fig:pps_spec}).

\begin{figure}[!ht]
	\centering
	\includegraphics[width=1\linewidth]{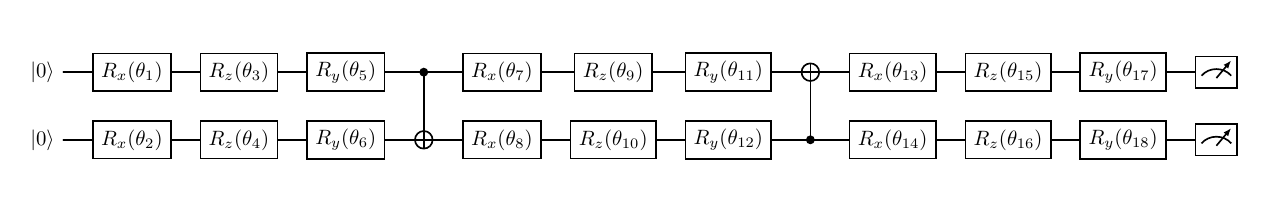}
	\caption{Parametrized circuit to implement the entire unitary operations in system of (CH$_3$O)$_2$PHO.}
	\label{fig:pqc_2bit}
\end{figure}

\textit{Gate Implementation:}
A parameterized two-qubit circuit is used to realize the sub-circuits shown in Figure~\ref{fig:exp_ghz} (dashed boxes). The parameterized quantum circuit contains single-qubit arbitrary angle rotation gate, rotation parameters and two-qubit gate (controlled-not gate,CNOT gate), which is shown in Fig.~\ref{fig:pqc_2bit}. We use this one as this is a fully compiled gate set.
Each parameter is optimized through the gradient descent algorithm. The duration of each single-qubit rotation gates is about $0.3~ms$, and the duration of CNOT gate is about $2~ms$. Their theoretical fidelity are both greater than $99\%$.

\textit{GHZ State Preparation --- } Finally, the circuits are executed concurrently on all nodes in the cluster. Preparing GHZ State across two neighboring nodes yields a state fidelity of $ 92.3\%$.

\section{Supplementary information for non-hermitian Hamiltonian Simulation}
\label{app:exp2}
Simulating non-Hermitian Hamiltonian dynamics-central to open quantum systems, $\mathcal{PT}$-symmetric optics, and gain-loss processes-remains a significant challenge in quantum science. Unlike conventional unitary evolution, non-Hermitian generators produce norm-non-preserving dynamics and complex spectra, rendering classical simulation especially demanding. Existing quantum algorithms for non-Hermitian channels often require complex ancillary structures. Here, we establish a thread-level parallel protocol, in which modular subtasks can be distributed across multiple quantum processors. This approach systematically enables scalable, parallel simulation of non-Hermitian dynamics, addressing a key bottleneck in both theory and experiment.

Many scientific and engineering tasks require simulating non-unitary evolution. Recent advances, such as linear combination of unitaries (LCU) and dilation methods, use ancillary qubits to enable non-Hermitian Hamiltonian simulation. As a representative protocol, the LCU framework approximates the time-evolution operator as a sum of unitaries:
\[
	e^{-iHt} \approx \sum_{k=0}^{K} \gamma_k U_k,
\]
with coefficients $\gamma_k$ and unitary circuits $U_k$ constructed by, e.g., Taylor or Chebyshev expansion.
For example, using the Taylor series expansion up to order $K$, we have
\[
	e^{-iHt} = \sum_{n=0}^{N} \frac{(-iHt)^n}{n!} + \mathcal{O}\left(\frac{(Ht)^{N+1}}{(N+1)!}\right).
\]
By substituting  $H = \sum_{m=1}^M \beta_m V_m$, where $\beta_m$ are coefficients and $V_m$ are unitary operators,
\[
	e^{-iHt} \approx \sum_{n=0}^{N} \frac{(-it)^n}{n!} \left(\sum_{m=1}^{M} \beta_m V_m\right)^n.
\]
This results in a form of $\sum_{k=0}^{K} \gamma_k U_k$, with
\begin{eqnarray}
	U_k&=& V_{m_1} V_{m_2} \dots V_{m_n},  \nonumber \\
	\gamma_k &=& \frac{(-it)^n}{n!} \beta_{m_1} \beta_{m_2} \dots \beta_{m_n}, \nonumber
\end{eqnarray}
where $m_1, m_2\dots m_n \in [1, M]$, and $K=(M^{N+1}-1)/(M-1)$.

For time-dependent or non-Hermitian Hamiltonians, a linear combination of Hamiltonian simulation (LCHS) protocol (see Algorithm~\ref{algo_hs}) is employed~\cite{an2023linear, Huo2023errorresilientmonte}, which naturally decomposes the evolution into independent unitary tasks suitable for parallel execution across QPU nodes. The method also needs ancilla systems, in our setting, we use cluster QPU, which leads a plug-and-play paradigm.
\begin{algorithm}[H]
	\caption{Hamiltonian Simulation for Non-Hermitian Dynamics}
	\label{algo_hs}
	\hspace{\algorithmicindent}
	\textbf{Input}: $A(t) = H(t) + iL(t)$, initial state $u_0$, simulation time $T$, precision epsilon.

	\hspace{\algorithmicindent}
	\textbf{Output}: Final state u(T)
	\begin{spacing}{1.03}
		\begin{algorithmic}[1]
			\State Decompose $A(t)$ into Hermitian part $H(t)$ and anti-Hermitian part $L(t)$.
			\State Define function $V(t, k) = T \exp(-i \int_0^t (H(s) + kL(s)) ds)$.
			\State Truncate the integral at $K = \frac{c}{\epsilon}$, where $c$ is a constant.
			\State Discretize the integral using a trapezoidal rule:
			\State $K = \frac{c}{\epsilon}$ and $M = \lfloor\frac{2K T}{\epsilon} \rfloor$
			\For{$j = 0$ to $M$}
			\State $k_j = -K + \frac{2jK}{M}$
			\State $w_j = (2 - (\text{if } j == 0 \text{ or } j == M)) \frac{K}{M}$
			\State $c_j = \frac{w_j}{\pi(1 + k_j^2)}$
			\State $U_j(t) = T \exp(-i \int_0^t (H(s) + k_j L(s)) ds)$
			\State $v_j = c_j U_j(T) u_0$
			\EndFor
			\State Sum up the contributions:
			\State $u(T) = \sum_{j=0}^M v_j$
			\State \Return $u(T)$
		\end{algorithmic}
	\end{spacing}
\end{algorithm}

\subsection{Non-Hermitian Hamiltonian}
We consider the open-system evolution of a single qubit governed by a time-independent non-Hermitian operator~\cite{bender2003must, wu2019observation}:
\begin{equation}
	\mathcal A(t) = H(t) - i L(t), \quad
	H(t) = \sigma_x, \quad
	L(t) = \openone + \sigma_z,
\end{equation}
where $\sigma_{x,y,z}$ denote the Pauli matrices and $\openone$ is the identity operator. The system is initialized in the ground state $\ket{\psi_0} = (1,0)$, and we measure the expectation values of $\sigma_{y}$ and $\sigma_{z}$.

This non-Hermitian evolution is experimentally realized on a cluster of $16\times2$-qubit nodes, with a total of over 1700 experimental runs (approximately 106 per node). The simulation protocol is based on the linear combination of Hamiltonian simulation (LCHS), which approximates the non-unitary evolution using a truncated Cauchy-integral formula, as detailed in Algorithm~\ref{algo_hs}. In addition, numerical simulations are performed using both this methodology and the exact integration of Schr{\"o}dinger's equation via a first-order Trotter step.

\begin{table}[!ht]
	\centering
	\begin{tabular}{@{}lll@{}}
		\toprule
		Symbol        & Meaning                   & Value                                \\
		\midrule
		$T$           & Total evolution time      & $0.1$-$1.0$ (step $0.1$)             \\
		$\Delta t$    & Trotter step size         & $0.01$                               \\
		$\varepsilon$ & LCHS quadrature step size & $0.2$                                \\
		$c$           & Constant in cutoff        & $0.5$                                \\
		$K$           & Integration cutoff        & $\lfloor c / \varepsilon\rfloor = 2$ \\
		$M$           & Number of nodes           & $\lfloor 2KT/\varepsilon\rfloor$     \\
		\bottomrule
	\end{tabular}
	\caption{Numerical parameters used in the simulation of non-Hermitian Hamiltonian evolution.}
	\label{tab:sim_params1}
\end{table}

\begin{figure}[!h]
	\centering
	\begin{tikzpicture}[thick]
		\ctikzset{scale=2}
		\tikzstyle{every node}=[font=\normalsize,scale=1]
		\tikzstyle{operator} = [draw,shape=rectangle
		,fill=white,minimum width=1em, minimum height=1em]
		\tikzstyle{operator2} = [draw,shape=rectangle,fill=pink,minimum width=2.5em, minimum height=2.5em]
		\tikzstyle{operator22} = [draw,shape=rectangle,fill=white,minimum width=3em, minimum height=9.5em]
		\tikzstyle{operator3} = [draw,shape=rectangle,fill=white,minimum width=3em, minimum height=1em]
		\tikzstyle{operator4} = [draw,shape=rectangle,dashed, minimum width=1.5cm, minimum height=1cm]
		\tikzstyle{operator5} = [draw,shape=rectangle,dashed, minimum width=7.5cm, minimum height=4cm]
		\tikzstyle{operator6} = [draw=pink,shape=rectangle,dashed, minimum width=5cm, minimum height=4cm]
		\tikzstyle{phase} = [fill,shape=circle,minimum size=3pt,inner sep=0pt]
		\tikzstyle{surround} = [fill=blue!10,thick,draw=black,rounded corners=2mm]
		\tikzstyle{ellipsis} = [fill,shape=circle,minimum size=2pt,inner sep=0pt]
		\tikzstyle{ellipsis2} = [fill=white,draw=black, shape=circle,minimum size=4pt,inner sep=0.5pt]
		\tikzset{meter/.append style={fill=white, draw, inner sep=5, rectangle, font=\vphantom{A}, minimum width=15,
					path picture={\draw[black] ([shift={(.05,.2)}]path picture bounding box.south west) to[bend left=40] ([shift={(-.05,.2)}]path picture bounding box.south east);\draw[black,-latex] ([shift={(0,.15)}]path picture bounding box.south) -- ([shift={(.15,-.08)}]path picture bounding box.north);}}}
		\node at (0.,0) (qin){$\ket{0}$};
		\node at (0.,-1) (qin){$\ket{0}$};
		\node at (0.1,0) (q2){};
		\node[meter] (end3) at (2.9,0) {} edge [-] (q2);
		\node at (0.1,-1) (q3) {};
		\node[] (end3) at (2.9,-1) {} edge [<-] (q3);
		\node[operator] (op22) at (0.5,-0) {$H$} ;
		\node[ellipsis2] (cc1) at (1,-0) {};
		\node[operator] (op22) at (1,-1) {$U_i$} edge [-] (cc1);
		\node[phase] (cc1) at (1.5,-0) {};
		\node[operator] (op22) at (1.5,-1) {$U_j$} edge [-] (cc1);
		\node[phase] (cc1) at (2.25,-0) {};
		\node[operator2] (op22) at (2.25,-1) {} edge [-] (cc1);
		\node at (2.25,-1) (qtex){\textcolor{black}{$M$}};
	\end{tikzpicture}
	\caption{Ancilla-assisted quantum circuit for measuring overlaps $\langle U_j^\dagger M U_i \rangle$ required in the LCHS protocol. By varying $M$, we obtain the desired real and imaginary parts of the relevant observables.}
	\label{fig:exp_nhs}
\end{figure}
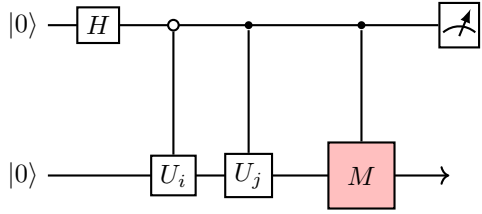

Table~\ref{tab:sim_params1} summarizes the numerical parameters. Here, $M$ is the number of nodes and thus sets the count of decomposed unitary terms. Employing the single-ancilla estimator, illustrated in Fig.~\ref{fig:exp_nhs}, each value of $T$ requires $1+\lfloor2KT/\varepsilon\rfloor$ individual experiments per observable. We sweep
\[
	T \in \{0.1,\,0.2, \dots,\,1.0\},
\]
and for each $T$ record the expectation of a random Hermitian operator $R$ as well as of $\sigma_x$ and $\sigma_z$. Altogether, measuring any one observable entails
\[
	\sum_{j=1}^{10}(1+\lfloor2KT_j/\varepsilon\rfloor)^2
\]
distinct experiments. Although the ideal count is $1770$, stepwise truncation in the computation of $M$ leads to a slight discrepancy, $1747$. The reason is one calculating the $T=0.6$ case. The truncation on machine made the $M=11$, which makes little influence.

The parameters above are selected via considering the efficiency of experiments and accuracy.
Throughout this protocol, the fidelity between the Trotter approximation and the LCHS implementation (with $\Delta t = 0.01$) mostly exceeds 99.2\%, confirming the accuracy of our linear-combination approach.
\begin{table}[ht]
	\centering
	\caption{Summary of $T$, nodes $M$, and resulting fidelity for each sampling point.}
	\label{tab:process_data1}
	\begin{tabular}{c|rrrrrrrrrr}
		\hline
		$k$      & 1      & 2      & 3      & 4      & 5      & 6      & 7      & 8      & 9      & 10     \\
		\hline
		$T$      & 0.1    & 0.2    & 0.3    & 0.4    & 0.5    & 0.6    & 0.7    & 0.8    & 0.9    & 1.0    \\
		$M$      & 2      & 4      & 6      & 8      & 10     & 11     & 14     & 16     & 18     & 20     \\
		Fidelity & 0.9999 & 0.9987 & 0.9945 & 0.9870 & 0.9808 & 0.9828 & 0.9929 & 0.9999 & 0.9964 & 0.9871 \\
		\hline
	\end{tabular}
\end{table}

In Table~\ref{tab:sigma_values}, we report the deviations between the experimentally measured expectation values of $\sigma_{y}$ and $\sigma_{z}$ and their numerically simulated counterparts, as well as those for a randomly generated Hermitian observable $R$, over the course of a non-Hermitian Hamiltonian evolution. The mean absolute deviation of $0.116$, quoted in the main text, is obtained by averaging all of the absolute values listed in this table.

\begin{table}[ht]
	\centering
	\caption{Deviations between experimental measurements and numerical simulations for $\langle\sigma_{y}\rangle$, $\langle\sigma_{z}\rangle$, and a random Hermitian observable $R$ during non-Hermitian Hamiltonian evolution.}
	\label{tab:sigma_values}
	\begin{tabular}{c|rrrrrrrrrr}
		\hline
		           & 1         & 2         & 3         & 4         & 5 & 6 & 7 & 8 & 9 & 10 \\
		\hline
		$\sigma_y$ &
		$-0.1208$  & $-0.1901$ & $-0.1898$ & $-0.0930$ & $-0.0456$ &
		$-0.0188$  & $ 0.1648$ & $ 0.2057$ & $ 0.2123$ & $ 0.1044$                          \\
		$\sigma_z$ &
		$-0.0497$  & $ 0.0891$ & $ 0.1104$ & $ 0.1780$ & $ 0.2054$ &
		$ 0.2645$  & $ 0.2498$ & $ 0.1037$ & $ 0.0895$ & $-0.0411$                          \\
		$R$        &
		$ 0.0167$  & $-0.0563$ & $-0.1193$ & $-0.0824$ & $-0.1006$ &
		$-0.1625$  & $-0.1040$ & $-0.0245$ & $-0.0385$ & $ 0.0471$                          \\
		\hline
	\end{tabular}
\end{table}

\subsection{Imaginary time evolution}
We further investigate the ground-state energy problem for a single-qubit Hamiltonian parameterized by a real coupling~$\gamma \in [0,2]$,
\begin{equation}
	H(\gamma) = 2\,\openone + \gamma\,\sigma_x,
	\label{eq:H-gamma2}
\end{equation}
where $\openone$ is the identity and $\sigma_x$ is the Pauli-$x$ operator. The analytic ground-state energy is $E_0(\gamma) = 2 - |\gamma|$, which serves as an independent benchmark. The ground state can also be found via imaginary-time evolution: the qubit is initialized in $\ket{\psi_0} = (1,0)^T$, propagated in a non-unitary fashion, and the ground-state energy is estimated as $E(\tau) = \langle \psi(\tau) | H(\gamma) | \psi(\tau) \rangle$.

Experimentally, this imaginary-time evolution is implemented as a linear combination of unitary evolutions via the truncated Cauchy-integral formula. On the cluster of $16\times2$-qubit nodes, imaginary-time evolution of $H(\gamma)$ for a total time $T=0.5$ is performed with more than 1331 experimental runs for obtaining single observable(about 83 per node). In addition to the ground-state energy, the expectation values of $\sigma_x$ and $\sigma_z$ are also measured for the resulting ground states.

To validate our experimental results, we benchmark three numerical simulations. First is the exact imaginary-time Trotter integration, in which the state is evolved by repeated application of $\exp[-H(\gamma)\Delta t]$ with a fixed step size $\Delta t = 0.01$ and $t = 0.5$.
The second is the LCHS-imaginary protocol, where the non-unitary propagator $\exp[-H(\gamma)\tau]$ is approximated via a truncated Cauchy integral with quadrature step $\varepsilon = 0.3$, shown in Algorithm~\ref{algo_hs}.
The third is the exact diagonalization on $H(\gamma)$, which yields the ground-state energy, to get a more approximated wavefunction, we use numerical imaginary-time evolution with $T=1.5$.

\begin{table}[!ht]
	\centering
	\begin{tabular}{@{}lll@{}}
		\toprule
		Symbol        & Meaning                  & Value                               \\
		\midrule
		$\gamma$      & parameter in Hamiltonian & $0.0$-$2.0$ (step $0.2$)            \\
		$T$           & total evolution time     & $0.5$                               \\
		$\Delta t$    & Trotter step             & $0.01$                              \\
		$\varepsilon$ & LCHS quadrature step     & $0.3$                               \\
		$c$           & a constant               & $1$                                 \\
		$K$           & integration cutoff       & $\lceil c/\varepsilon\rceil = 3$    \\
		$M$           & number of nodes          & $\lfloor 2KT/\varepsilon\rfloor=10$ \\
		\bottomrule
	\end{tabular}
	\caption{Numerical parameters used in all simulation of imaginary-time evolution.}
	\label{tab:sim_params2}
\end{table}
The simulation parameters are summarized in Table~\ref{tab:sim_params2}. In particular, $M$ denotes the number of nodes, which in turn determines the total number of decomposed unitary terms, $U_i, U_j$.
To extract the required expectation values, we design a single-ancilla estimator, illustrated in Fig.~\ref{fig:exp_nhs}. Thus, this decomposition requires $11\times11$ ($M+1\times M+1$) individual experiments. We sweep the parameter $\gamma$ across eleven uniformly spaced values,
\[
	\gamma_k = 0.2\,k,\quad k = 0,1,\dots,10,
\]
and for each $\gamma_k$ we record the expectation values of the Hamiltonian $H(\gamma)$ as well as of $\sigma_x$ and $\sigma_z$ on the corresponding ground state. Consequently, measuring any single observable entails
\[
	11 \times 11 \times 11 = 1331
\]
distinct experiments.

At the conclusion of the imaginary-time evolution ($T = 0.5$), the fidelity between the Trotter approximation and the LCHS protocol exceeds $99.5 \%$ for most sampling points, thereby confirming the accuracy of the linear-combination approach. Detailed fidelity values are listed in Table~\ref{tab:processed_values2}.
\begin{table}[ht]
	\centering
	\caption{Fidelity between the Trotter method and the LCHS protocol after imaginary-time evolution ($T = 0.5$) for $\gamma_k = 0.2\,k$ (indices $k=1,\dots,10$).}
	\label{tab:processed_values2}
	\begin{tabular}{c|rrrrrrrrrr}
		\hline
		         & 1        & 2        & 3         & 4         & 5 & 6 & 7 & 8 & 9 & 10 \\
		\hline
		fidelity &
		$0.9990$ & $0.9969$ & $0.9950$ & $0.9944$  & $0.9952$  &
		$0.9967$ & $0.9983$ & $0.9995$ & $0.99998$ & $0.99977$                          \\
		\hline
	\end{tabular}
\end{table}

In Table~\ref{tab:sigma_values_2}, we report the deviations between the experimentally measured expectation values of $\langle\sigma_x\rangle$ and $\langle\sigma_z\rangle$ and their numerically simulated counterparts, as well as those for the Hamiltonian observable $H(\gamma)$, at the conclusion of an imaginary-time evolution ($T=0.5$). The overall mean absolute deviation of $0.136$, quoted in the main text, is obtained by averaging all of the absolute values listed in this table.

\begin{table}[ht]
	\centering
	\caption{Deviations between experimental measurements and numerical simulations for $\langle\sigma_x\rangle$, $\langle\sigma_z\rangle$, and the Hamiltonian $H(\gamma)$ after imaginary-time evolution ($T=0.5$). $\gamma$ is set as $0.2k$, where $k$ label this horizontal axis.}
	\label{tab:sigma_values_2}
	\begin{tabular}{c|*{11}{r}}
		\hline
		           & 1         & 2         & 3         & 4         & 5         & 6 & 7 & 8 & 9 & 10 & 11 \\
		\hline
		$\sigma_1$ &
		$-0.3055$  & $-0.2313$ & $-0.2008$ & $-0.3122$ & $-0.2248$ &
		$-0.2227$  & $-0.0779$ & $-0.0072$ & $-0.0351$ & $ 0.0761$ & $ 0.0498$                           \\
		$\sigma_2$ &
		$-0.2459$  & $-0.1625$ & $-0.0745$ & $-0.2681$ & $-0.1279$ &
		$-0.1384$  & $-0.0746$ & $-0.0652$ & $-0.0296$ & $-0.0692$ & $-0.1333$                           \\
		$\sigma_3$ &
		$-0.0108$  & $ 0.0705$ & $-0.0115$ & $ 0.0529$ & $ 0.1826$ &
		$ 0.1067$  & $ 0.1317$ & $ 0.1807$ & $ 0.1328$ & $ 0.1955$ & $ 0.3062$                           \\
		\hline
	\end{tabular}
\end{table}

Finally, we present the experimental implementation on hardware. This is carried out in a two-qubit cluster of the Gemini-Lab desktop spectrometer, as described in the Appendix~\ref{app:hardware}.

The overall procedure follows a structure similar to the quantum circuit cutting experiments detailed in Appendix~\ref{app:exp1}, and here we focus on a single representative thread.

As in previous sections, a relaxation-based initialization sequence is applied to convert the thermal equilibrium state into a pseudo-pure state that approximatesates $\ket{00}$. The resulting state fidelity reaches $99.5\%$.

Next, a parameterized two-qubit quantum circuit is used to implement the sub-circuits illustrated in Fig.~\ref{fig:exp_nhs}. The circuit consists of arbitrary angle single-qubit rotation gates and a two-qubit controlled-NOT (CNOT) gate. All parameters are optimized using a gradient descent algorithm. Each single-qubit rotation has a duration of approximately $0.3~\mathrm{ms}$, while the CNOT gate duration is about $2~\mathrm{ms}$. Theoretical fidelities for both types of gates exceed $99\%$.
Finally, the compiled circuit is executed and the measurement is performed to extract the desired observables.

\end{document}